\documentclass[pre,preprint,unsortedaddress]{revtex4-1} 

\usepackage{amsmath}  
\usepackage{amsfonts} 
\usepackage{graphicx} 
\usepackage{enumitem} 
\usepackage{xcolor} 
\usepackage[utf8]{inputenc} 
\usepackage{tikz} 
\usepackage{csquotes}
\usepackage[subrefformat=parens,labelformat=parens]{subfig}

\begin{document}


\title{Random spherical graphs}

\author{Alfonso Allen-Perkins}
\email{alfonso.allen@hotmail.com}
\affiliation{Complex System Group, Universidad Polit\'ecnica de Madrid, 28040-Madrid, Spain.\\
Instituto de F\'isica, Universidade Federal da Bahia, 40210-210 Salvador, Brazil.}%

\date{\today}

\begin{abstract}
This work addresses a modification of the random geometric graph (RGG) model by considering a set of points uniformly and independently distributed on the surface of a $(d-1)$-sphere with radius $r$ in a $d-$dimensional Euclidean space, instead of on an unit hypercube $[0,1]^d$ . Then, two vertices are connected by a link if their great circle distance is at most $s$. In the case of $d=3$, the topological properties of the random spherical graphs (RSGs) generated by this model are studied as a function of $s$. We obtain analytical expressions for the average degree, degree distribution, connectivity, average path length, diameter and clustering coefficient for RSGs. By setting $r=\sqrt{\pi}/(2\pi)$, we also show the differences between the topological properties of RSGs and those of two-dimensional RGGs and random rectangular graphs (RRGs). Surprisingly, in terms of the average clustering coefficient, RSGs look more similar to the analytical estimation for RGGs than RGGs themselves, when their boundary effects are considered. We support all our findings by computer simulations that corroborate the accuracy of the theoretical models proposed for RSGs.

\end{abstract}

\maketitle 

\section{Introduction}

Random geometric graphs (RGGs) models contain a collection of nodes distributed randomly throughout a given domain  of typically two or three dimensions, together with connecting links between pairs of nodes that exist with a probability related to the vertex locations \cite{Penrose03,Dall02,Franceschett2007,Walters11}. An intense theoretical research on RGGs has been triggered by the study of complex spatial networks \cite{Barthelemy11}, that is, networks that emerge under certain geometrical constraints. RGGs have been applied to a variety of complex systems including in city growth \cite{Watanabe10}, power grids \cite{Xiao11}, nanoscience \cite{Kyrylyuk11}, epidemiology \cite{Miller09,Danon11,Nekovee07,Isham11,Toroczkai07}, forest fires \cite{Pueyo10}, social networks \cite{Palla07,Parshani11,Wang09,Diaz09}, and wireless communications \cite{Gilbert1959,Haenggi09,Li09,Wang09,Estrin1999,Pottie2000,Gupta1999}, among others.

Most research on RGGs has been carried out in networks in which the nodes of the graph are distributed randomly and independently in a unit square and two nodes are connected if they are inside a disk of a given radius, centered at one of the nodes. Hereafter, the term RGG is reserved for that case. However, various modifications of the RGG model have been introduced to account for
more realistic scenarios. This includes the generalization of RGGs to random rectangular networks (RRGs) \cite{Estrada15a,Estrada15b,Estrada16a,Estrada16b,Arias18}, the use of a general connection function \cite{Dettmann16,Gilles16}, the embedding of the nodes into fractal domains \cite{Coon12} and into the two-dimensional unit torus \cite{Dettmann17}, or the use of low-discrepancy sequences \cite{Estrada17a} to mention a few. Likewise, different geometric rules for generating the edges have been studied. This is the case of the random proximity graphs and their generalization \cite{Jaromczyk92,Gabriel69}, the scale-free spatial networks \cite{Herrmann03,Jacob15}, or the Waxman model \cite{Waxman88}.

It has been shown that the geometrical details of the confined space boundary (such as corners, edges, faces or the length of the perimeter) dominate the topological properties of RGG and RRG models, and play a fundamental role on the dynamical properties of the resulting graphs \cite{Estrada15a,Estrada15b,Estrada16a,Estrada16b,Arias18,Coon12}. In this work, we develop a model that modifies the RGG and RRG approachs by allowing the embedding of the nodes on a sphere instead of a unit square or a unit rectangle. Our main goal is to investigate how the lack of borders of the graphs generated by the new model impacts some network-theoretic invariants that are commonly used to characterize the structural properties of networks, namely: the average degree, connectivity, degree distribution, average path length, diameter and clustering coefficient. Previous investigations on other domains without borders were restricted to spectral statistics \cite{Dettmann17}. We found analytical expressions and bounds for all the topological properties mentioned above, and provide computational evidence of the accuracy of these approaches for RSGs. 

The paper is organized as follows. In Sec.~\ref{MODEL}, we present the RSG model. The topological properties of RSGs are analyzed in Sec.~\ref{Properties}. Finally, our conclusions are summarized in Sec.~\ref{conclusions}.


\section{Definition of the model}
\label{MODEL}

A random spherical graph (RSG) is a network where each of its $N$ nodes is
assigned random coordinates on the $(d-1)$-sphere centered at the origin, in a $d-$dimensional Euclidean space:

\begin{eqnarray}
S^{d-1}(r)=\left \{ x \in \mathbb{R}^d:\left \| x \right \|=r \right \},
\label{def_unit_sphere}
\end{eqnarray}

\noindent  where $\left \| . \right \|$ is the standard Euclidean metric on $\mathbb{R}^d$. Then, two vertices are connected by a link if their \textit{great circle distance} (i.e., the shortest distance between two points, measured along the surface of the sphere) is at most $s$. Hereafter, we will consider only the case $d=3$, which corresponds to a sphere of radius $r$. In the case of $r=1$, we call our surface the \textit{unit ball}, $S^2$. Now, the RSG is defined by distributing uniformly and
independently $N$ vertices in the sphere and then
connecting two points $i$ and $j$ by an edge if their circle distance distance $\delta(i,j)$
is less than or equal to $s$, the \textit{connection radius} of the RSG. According to the haversine formula, $\delta(i,j)$ is described by

\begin{eqnarray}
\delta(i,j)=2r\arcsin\left ( \sqrt{\sin^2\left ( \frac{\phi_j -\phi_i }{2} \right ) +\cos\left ( \phi_i \right )\cos\left ( \phi_j \right )\sin^2\left ( \frac{\lambda_j -\lambda_i }{2} \right ) } \right )
\label{haversine}
\end{eqnarray}

\noindent where $\phi_P$ and $\lambda_P$ are the latitude and longitude of point $P$ (in radians), respectively. The conversion from the geographic coordinates $(\phi _P,\lambda _P)$ to a cartesian vector $(x_P,y_P,z_P)$ of a position P is given by:

\begin{eqnarray}
x_P = r\cos(\phi _P) \cos(\lambda _P),\nonumber \\
y_P = r\cos(\phi _P) \sin(\lambda _P),\\
z_P = r\sin(\lambda _P). \nonumber
\label{conversion}
\end{eqnarray}

\begin{figure}[h!]
\centering
\subfloat[]{\label{cap}\includegraphics[width=0.5\textwidth]{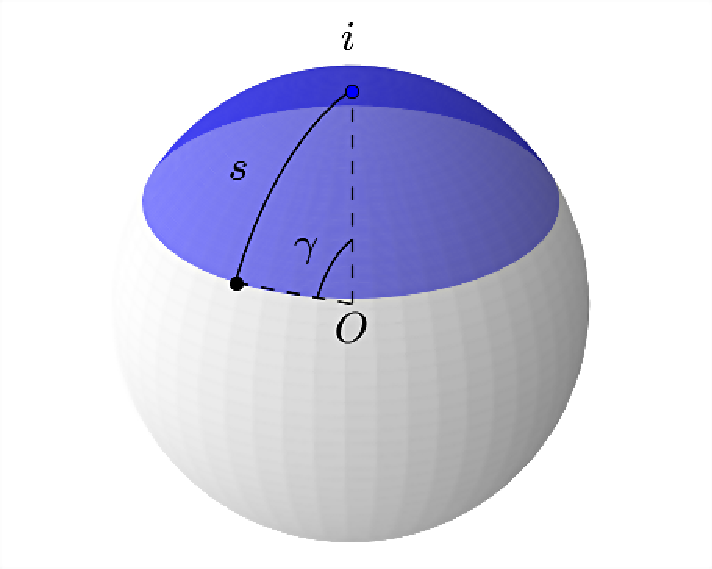}}
\subfloat[]{\label{example_rsg}\includegraphics[width=0.51\textwidth]{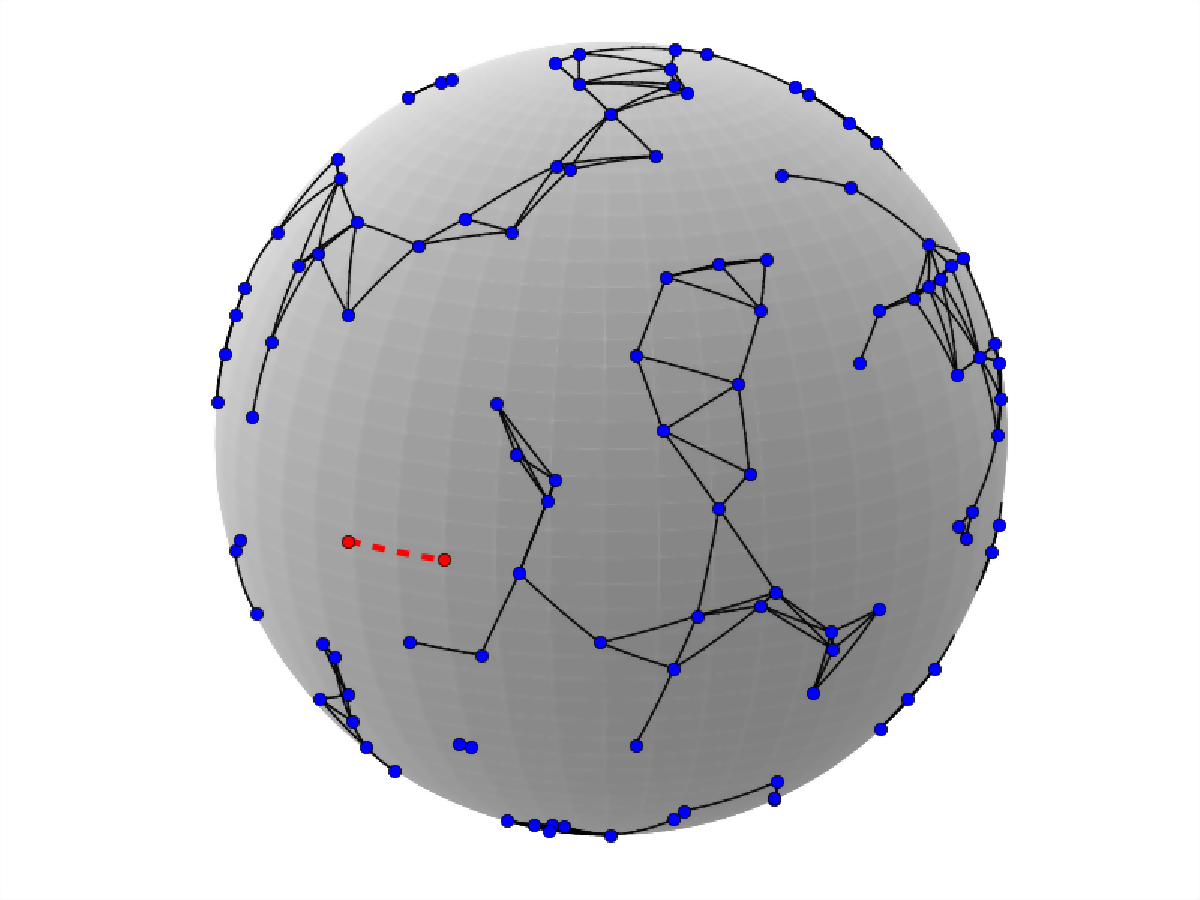}}
\caption{(a) Example of a spherical cap $SC(i,s,r)$ (blue surface) with polar angle $\gamma$. All points $j$ in $SC(i,s,r)$ satisfy the condition $\delta(i,j)\leq s$. (b) A random spherical graph with $N=200$ nodes and $s=0.27$, when $r=1$. A red dashed line of arc length $s=0.27$ is included as a point of reference.}
\label{model}
\end{figure}

Let us now consider a node $i$ and the set of points in the sphere where $\delta(i,j)\leq s$. As can be observed in Fig.~\subref*{cap}, these points define a spherical cap \enquote{centered} in $i$, denoted as $SC(i,s,r)$. Note that, in the case of $S^2$ (i.e. $r=1$), the great circle distance $s$ and the polar angle of $SC(i,s,1)$, denoted as $\gamma=s/r$, are equivalent. Thus, according to our model, we connect every node to the other vertices that lie inside its cap. For this reason, the surface of the cap represents the \textit{connection area} of a given point in the sphere. In Fig.~\subref*{example_rsg} we show an example of RSG.

It is easy to see that our model has some similarities with the RGG \cite{Penrose03,Dall02} and RRG \cite{Estrada15a} graphs. In all the cases, $N$ nodes are distributed uniformly and independently in a finite surface, and two points are connected by an edge if the shortest distance between them, measured along the surface, is less than or equal to a given threshold. However, in the case of RSGs, an important difference has been introduced: the unit ball has no borders, while the other types of random graphs do. As we will see in the next sections of this paper, the lack of borders and the spherical symmetry of RSGs influence their topological parameters and also facilitate their analytical study. 


\section{Topological properties of RSG\MakeLowercase{s}}
\label{Properties}

\subsection{Average degree}
\label{Degree}

Following Ref.~\cite{Estrada15a}, we start the study of the topological properties of RSGs by
considering an analytical expression for the average degree of the network $\bar{k}=M/N$, where $N$ is the number of nodes and $M$ is the amount of edges. As we will see in the next sections, this network-theoretic invariant is fundamental to understand other topological parameters of RSGs, as well as their relations with those of RGGs and RRGs.

Let $k_i$ be the degree of a vertex $i$ (i.e. the number of edges incident to $i$) and let us consider that, for a given node $i$, there are $N-1$ nodes distributed in the rest of the sphere. Since the nodes are uniformly and independently
distributed, the expected degree of a node $i$ is

\begin{eqnarray}
\mathbf{E}\left [ k_i (s) \right ] = (N-1)\frac{A_{SC_i}}{4\pi r^2},
\end{eqnarray}

\noindent where $r$ is the radius of the sphere ($r=1$ for $S^2$), and $A_{SC_i}$ is the surface of $SC(i,s,r)$, given by

\begin{eqnarray}
A_{SC_i}=\left\{\begin{matrix}
0\leq s\leq \pi r & 2\pi r^2\left ( 1-\cos\left ( \frac{s}{r} \right ) \right ) \\
\pi r < s  & 4\pi r^2
\end{matrix}\right..
\label{area_cap}
\end{eqnarray}

\noindent Note that $SC(i,s,r)$ is equal to the area of the sphere, when $s  \geq \pi r$. Considering that for a given value of $s$ all the points in $SC(i,s,r)$ are within the sphere, it is possible to see that $A_{SC_i}=A_{SC_j}$ for $i\neq j$. Therefore, the expected degree of each node is the same and, consequently, its value is equal to the expected average degree of the RSG (see Eq.~\ref{eq_grado_medio}).

\begin{eqnarray}
\mathbf{E}\left [ \bar{k}(s) \right ]=\frac{\int_{S^2} \mathbf{E}\left [ k_i(s) \right ] \mathrm{d} S}{4\pi r^2}=\mathbf{E}\left [  k_i(s) \right ]\frac{\int_{S^2} \mathrm{d}S}{4\pi r^2}=\mathbf{E}\left [  k_i(s) \right ]=\frac{N-1}{2}\left ( 1-\cos\left ( \frac{s}{r} \right ) \right )
\label{eq_grado_medio}
\end{eqnarray}

\begin{figure}[h!]
\centering
\includegraphics[width=0.5\textwidth]{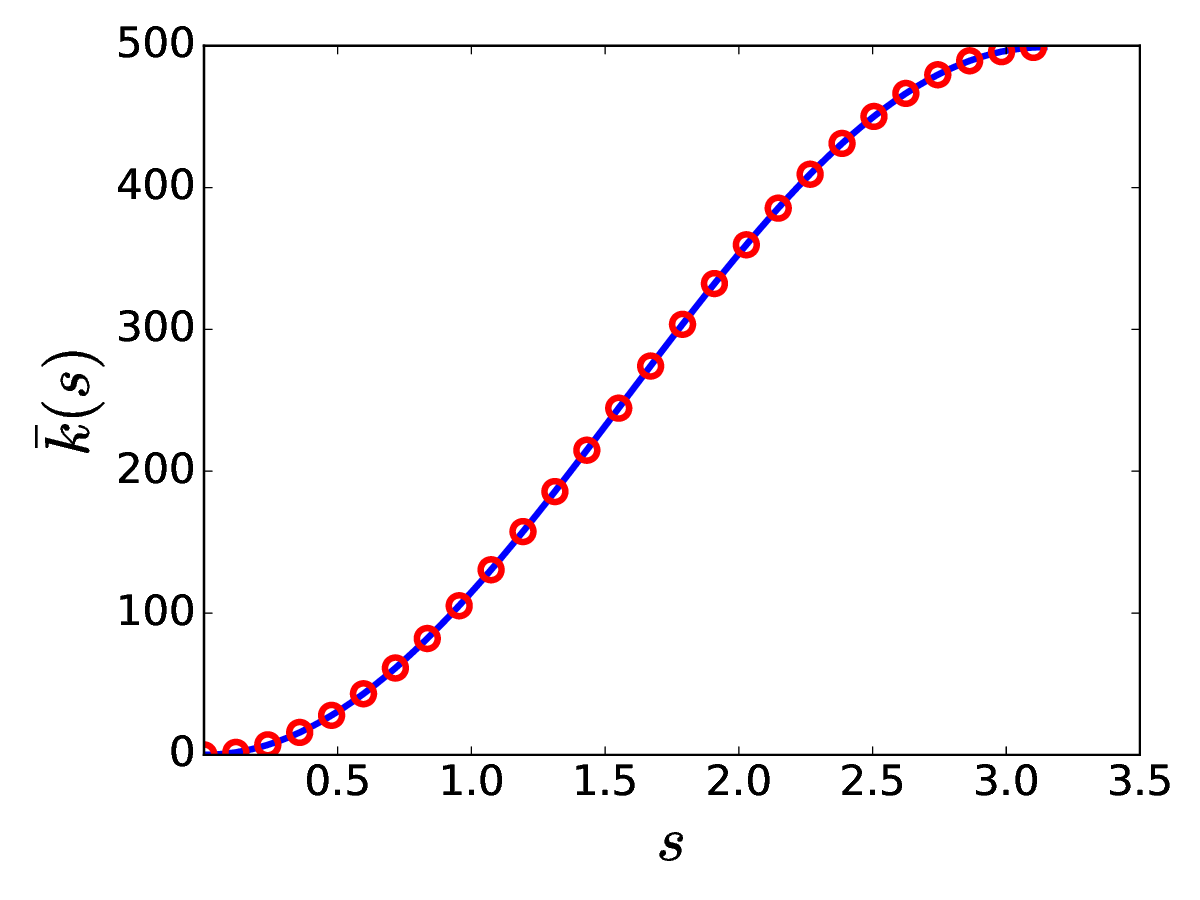}
\caption{Dependence of the average degree $\bar{k}$ on $s$ for a RSG with $N=500$ nodes, when $r=1$. The symbols indicate the numerical results (averaged over 100 realizations), and the blue continuous line corresponds to the analytical prediction of Eq.~\ref{eq_grado_medio}.}
\label{grado_medio}
\end{figure}

Fig.~\ref{grado_medio} illustrates the dependence of the average degree $\bar{k}(s)$ on the connection radius $s$, for the unit ball (i.e. $r=1$). As can be observed, there is a complete agreement between the results of Eq.~\ref{eq_grado_medio} and those obtained from computer simulations.

To study the influence of the borders on the expected average degree of RSGs, we compare our analytical findings with those for RGGs and RRGs. The expected average degree for RRGs with $N$ nodes embedded in a rectangle with sides of lengths $a$ and $b$ (where $b \leq a$ and $b=1/a$) was obtained analytically by Estrada and Sheerin in Ref.~\cite{Estrada15a}, and is given by

\begin{eqnarray}
\mathbf{E}\left [ \bar{k}(s) \right ]=\frac{N-1}{ab} \left \langle S_{RRG} \right \rangle =\frac{N-1}{(ab)^2}f(s,a)=(N-1)f(s,a),
\label{eq_grado_medio_rectangulo}
\end{eqnarray}

\noindent where $s$ is the connection radius (i.e. the maximum shortest distance between two connected nodes of the RRG), $\left \langle S_{RRG} \right \rangle = f(s,a)/(ab)=f(s,a)$ represents the average \textit{connection area} of the rectangle (i.e. the average area within radius $s$ of a point which lies in the rectangle), and

\begin{eqnarray}
f(s,a)=\left\{\begin{matrix}
0\leq s \leq b & \pi s^2 ab -\frac{4}{3}(a+b)s^3+\frac{1}{2}s^4\\  
b\leq s \leq a & -\frac{4}{3}a s^3-s^2b^2+\frac{1}{6}b^4+a\left ( \frac{4}{3}s^2+\frac{2}{3}b^2 \right )\sqrt{s^2-b^2}+\\
 & 2s^2\arcsin\left ( \frac{b}{s} \right )\\  
a\leq s \leq \sqrt{a^2+b^2} & -s^2(a^2+b^2)+\frac{1}{6}(a^4+b^4)-\frac{1}{2}s^4\\
& +b\left ( \frac{4}{3}s^2+\frac{2}{3}b^2 \right )\sqrt{s^2-a^2}+a\left ( \frac{4}{3}s^2+\frac{2}{3}b^2 \right )\sqrt{s^2-b^2}\\
& -2abs^2\left ( \arccos\left ( \frac{b}{s} \right )-\arcsin\left ( \frac{a}{s} \right ) \right )
\end{matrix}\right ..
\label{con_area_rect}
\end{eqnarray}

\begin{figure}[h!]
\centering
\subfloat[]{\label{comparacion_grados_medios}\includegraphics[width=0.5\textwidth]{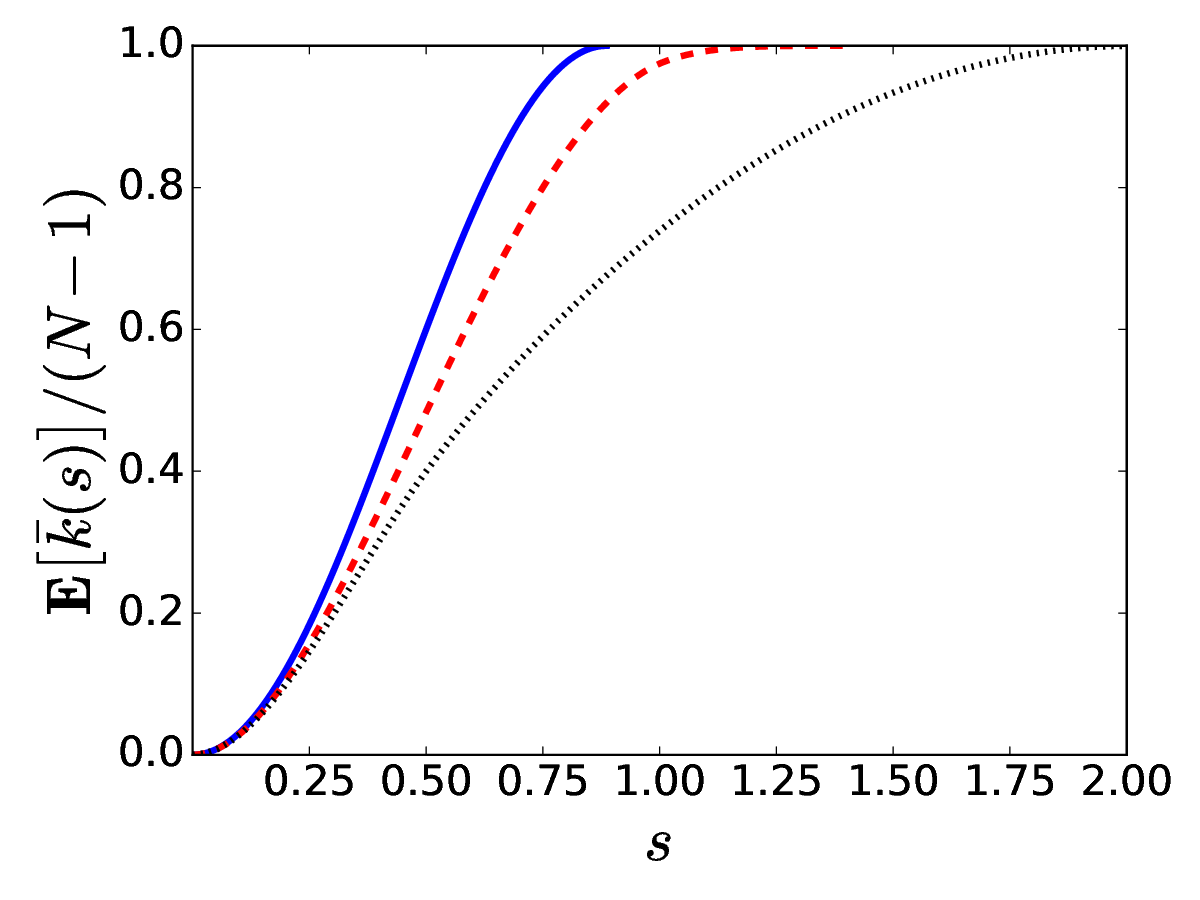}}
\subfloat[]{\label{connection_areas}\includegraphics[width=0.5\textwidth]{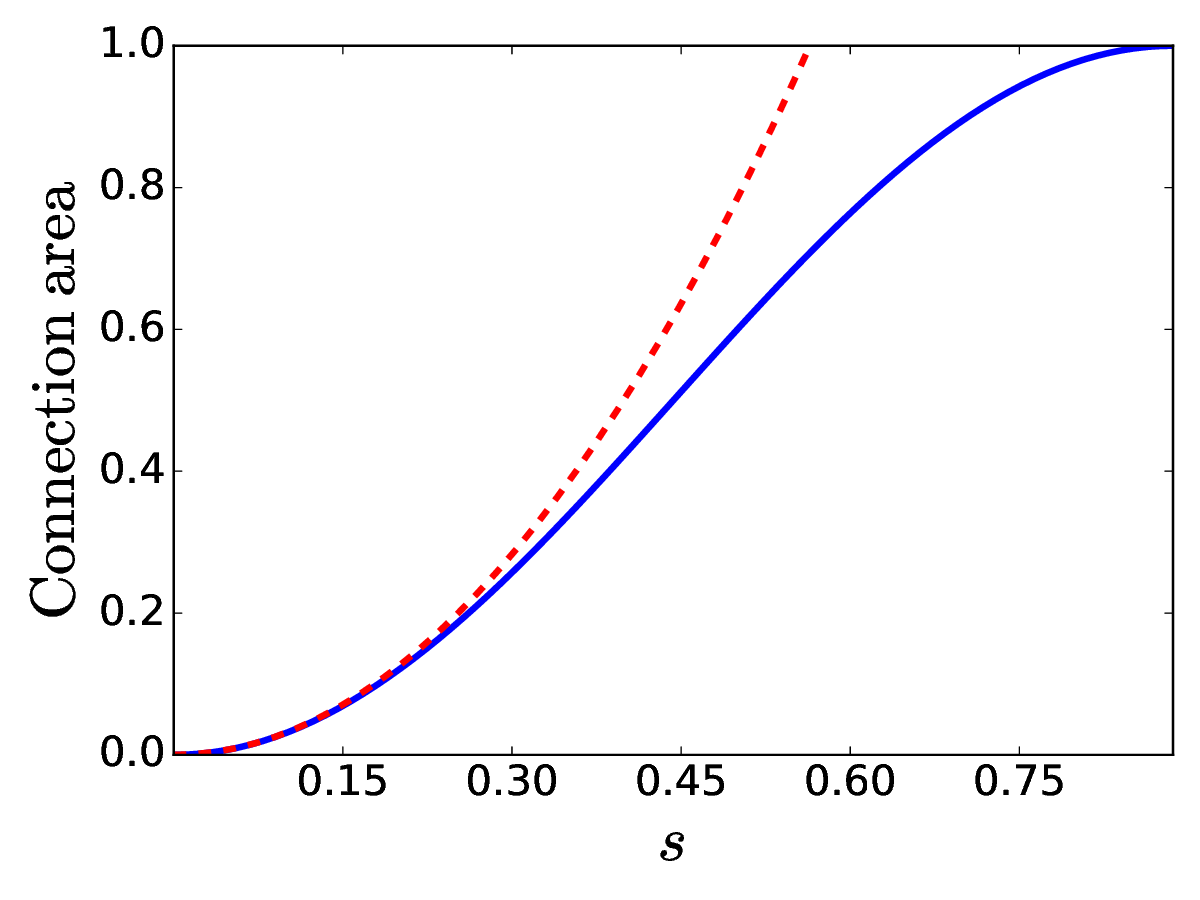}}
\caption{(a) Dependence of the expected (normalized) average degree $\mathbf{E}\left [  \bar{k}(s) \right ]/ (N-1)$ on $s$ for RSGs with radius $r=\sqrt{\pi}/(2\pi)$ (blue continuous line, Eq.~\ref{eq_grado_medio}), RGGs (red dashed line, Eq.~\ref{eq_grado_medio_rectangulo}) and RRGs with $a=2$ and $b=1/2$ (black dotted line, Eq.~\ref{eq_grado_medio_rectangulo}). (b) Dependence of the connection area on $s$ for RSGs with radius $r=\sqrt{\pi}/(2\pi)$ (blue continuous line, Eq.~\ref{area_cap}), and RGGs and RRGs (red dashed line, $S_{RRG}=\pi s^2$).}
\label{borders_comp}
\end{figure}

\noindent Then, we define a sphere that has the same area as the unit square (i.e. the radius of the new sphere is $r=\sqrt{\pi}/(2\pi)$ units). In this way, the uniform and independent distribution of nodes in each system takes place in the same area. In Fig.~\subref*{comparacion_grados_medios} we show the dependence of the expected average degree on $s$ for RSGs (with radius $r=\sqrt{\pi}/(2\pi)$), RGGs (with $a=1$ and $b=1$), and RRGs (with $a=2$ and $b=1/2$). As can be observed, for a given value of $s>0$ and $N$, the largest (smallest) expected average degree is obtained for RSGs (RRGs). This result is apparently surprising given that the connection areas of the RGGs and RRGs are larger than those of the RSGs, when $s>0$ and $r=\sqrt{\pi}/(2\pi)$ (see Fig.~\subref*{connection_areas}). The main reason for this behavior is the lack of borders of the RSGs. As indicated before, the whole area within the connection radius $s$ of a point lies in the surface of the sphere. However, in the case of RGGs and RRGs, their boundaries imply that there are points whose connection areas lie partially outside the borders of the square or rectangle, respectively. Those nodes have an expected smaller degree. Therefore, they reduced the average connection area and the expected average degree of their systems (see Fig.~\subref*{connection_areas}). As our data show, the larger the length of the perimeter, the smaller the expected average degree.


Finally, we compare our analytical findings for RSGs with those for RGGs and RRGs that have periodic boundaries (PB-RGGs and PB-RRGs, respectively). Unlike the case of the standard RGGs and RRGs, these systems (i.e., RSGs, PB-RGGs and PB-RRGs) satisfy the condition that the whole connection area of a point lies in the surface of the model considered. Consequently, in these graph models there are no nodes that could reduce the average connection area (i.e., the connection area of all the points is the same) and the expected average degree.

In the case of PB-RRGs with $N$ nodes embedded in a rectangle with sides of lengths $a$ and $b$ (where $b \leq a$ and $b=1/a$), the expected average degree is

\begin{eqnarray}
\mathbf{E}\left [ \bar{k}(s) \right ]=(N-1) f^{PB}(s,a),
\label{eq_grado_medio_rectangulo_PB}
\end{eqnarray}

\noindent where $s$ is the connection radius, and $f^{PB}(s,a)$ represents the (average) connection area of the PB-RRGs, given by

\begin{eqnarray}
f^{PB}(s,a)=
\left\{\begin{matrix}
0\leq s \leq \frac{1}{2}b & \pi s^2\\  
\frac{1}{2}b\leq s \leq \frac{1}{2}a & bs\sqrt{1-\frac{b^2}{4s^2}}+2s^2\arccos\left ( \sqrt{1-\frac{b^2}{4s^2}}\right )\\  
\frac{1}{2}a\leq s \leq \sqrt{\frac{1}{4}a^2+\frac{1}{4}b^2} & bs\sqrt{1-\frac{b^2}{4s^2}}+2s^2\arccos\left ( \sqrt{1-\frac{b^2}{4s^2}}\right )\\
& +as\sqrt{1-\frac{b^2}{4s^2}}-2s^2\arccos\left ( \sqrt{1-\frac{a^2}{4s^2}}\right )\\
\sqrt{\frac{1}{4}a^2+\frac{1}{4}b^2} \leq s  & ab
\end{matrix}\right .,
\label{connection_area_rectangulo_PB}
\end{eqnarray}

\begin{figure}[h!]
\centering
\subfloat[]{
   \begin{tikzpicture}
        \node[anchor=south west,inner sep=0] (image) at (0,0) {\includegraphics[width=0.5\textwidth]{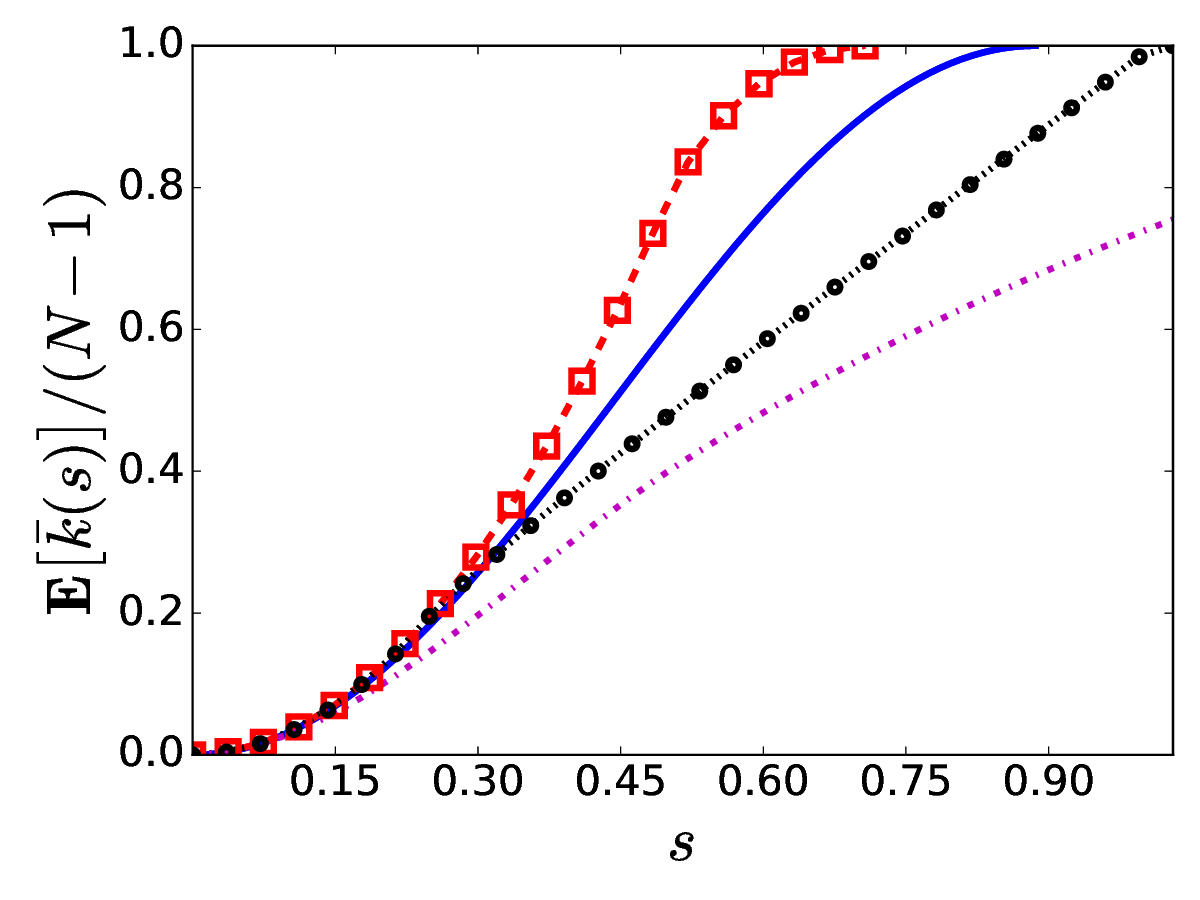}};
        \begin{scope}[x={(image.south east)},y={(image.north west)}]
            \node[anchor=south west,inner sep=0] (image) at (0.61,0.17) {\includegraphics[width=0.177\textwidth]{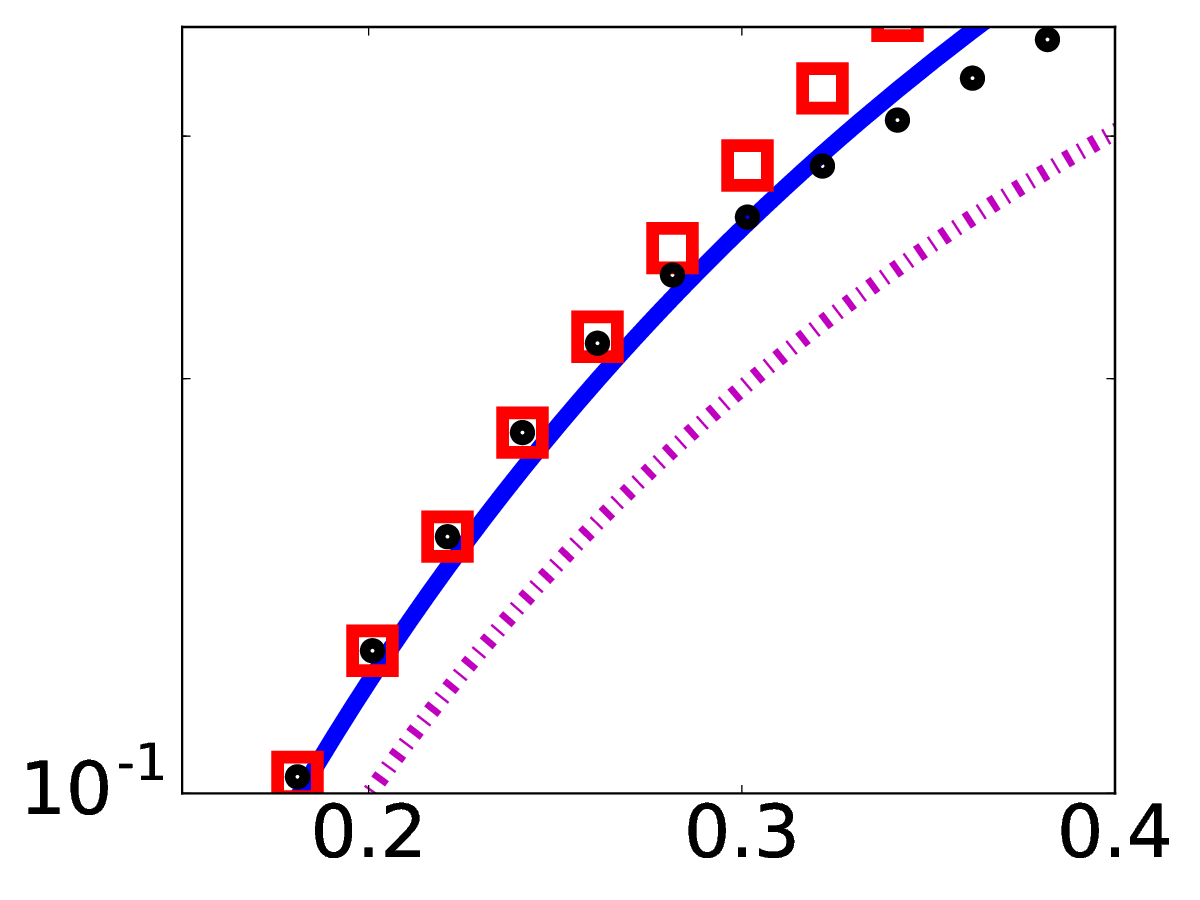}};
        \end{scope}
    \end{tikzpicture}
\label{comparacion_grados_medios_PERIODIC}
}
\subfloat[]{\label{otra_nueva_fig}\includegraphics[width=0.5\textwidth]{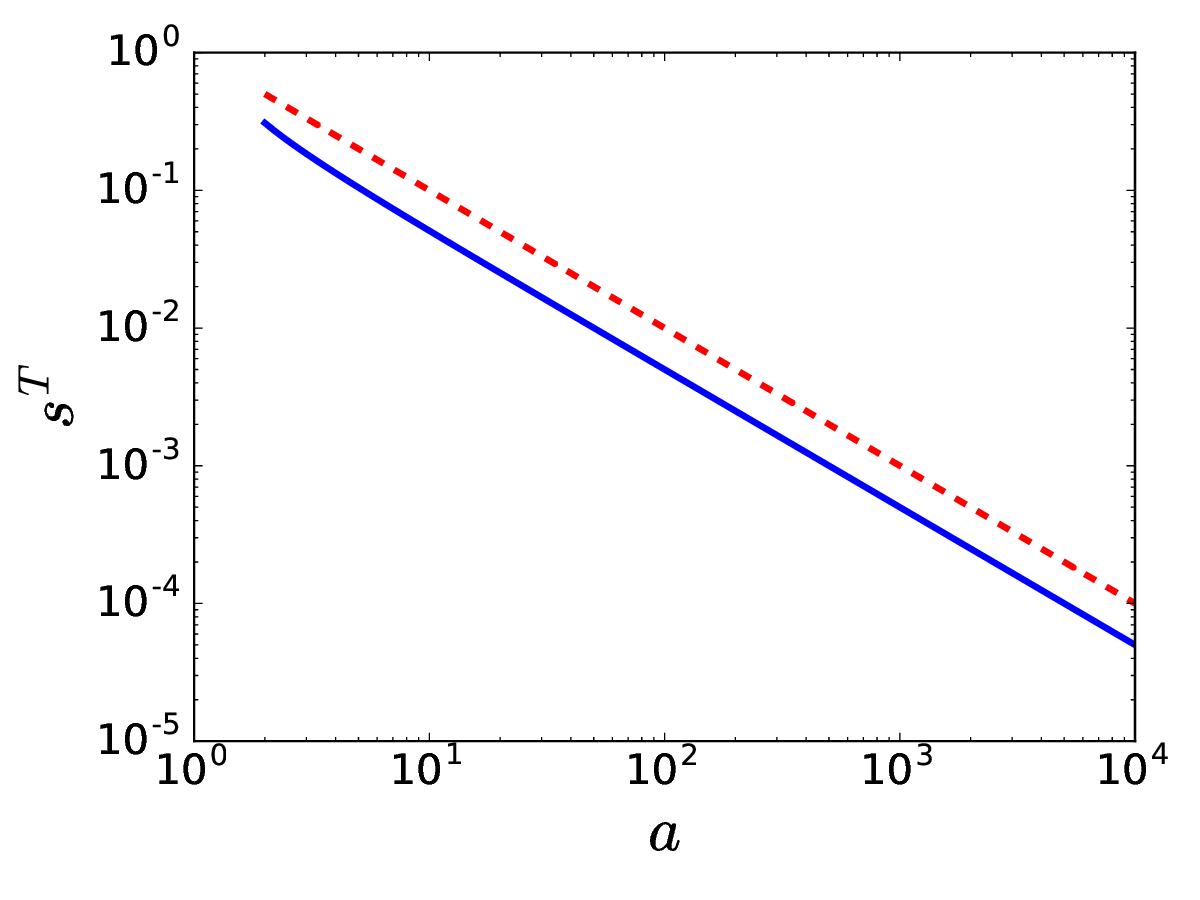}}
\caption{Dependence of the expected (normalized) average degree $\mathbf{E}\left [  \bar{k}(s) \right ]/ (N-1)$ on $s$ for RSGs with radius $r=\sqrt{\pi}/(2\pi)$ (blue continuous line, Eq.~\ref{eq_grado_medio}), PB-RGGs (red dashed line, Eq.~\ref{eq_grado_medio_rectangulo_PB}), PB-RRGs with $a=2$ and $b=1/2$ (black dotted line, Eq.~\ref{eq_grado_medio_rectangulo_PB}), and RRGs with $a=2$ and $b=1/2$ (magenta dash-dotted line, Eq.~\ref{eq_grado_medio_rectangulo}). The symbols indicate numerical results (averaged over 100 realizations) for PB-RGGs (red squares), and PB-RRGs with $a=2$ and $b=1/2$ (black dots). The inset shows (on a semi-logarithmic scale) greater detail of the region where the expected (normalized) average degree of the PB-RRGs (black dots) becomes smaller than the value of the RSGs (blue line). (b) Dependence of the threshold connection radius $s^T$ (blue continuous line, Eq.~\ref{cal_sT}) on the length of the PB-RRGs' longest side, $a$. The red dashed line is a guide for the eye proportional to $1/a$.}
\label{grado_periodico}
\end{figure}

\noindent (see Appendix for the derivation of Eqs.~\ref{eq_grado_medio_rectangulo_PB} and \ref{connection_area_rectangulo_PB}). As expected, according to Eqs.~\ref{con_area_rect} and \ref{connection_area_rectangulo_PB}, the average connection area of PR-RGGs (PR-RRGs) is larger than those of RGGs (RRGs, see Fig.~\subref*{comparacion_grados_medios_PERIODIC}).

In Fig.~\subref*{comparacion_grados_medios_PERIODIC} we show the dependence of the expected (normalized) average degree on $s$ for RSGs (with radius $r=\sqrt{\pi}/(2\pi)$), PB-RGGs (with $a=1$ and $b=1$), and PB-RRGs (with $a=2$ and $b=1/2$). We also exhibit the computer results (averaged over 100 realizations) for the expected (normalized) average degree of PB-RGGs and PB-RRGs. As can be observed, computer calculations are in good agreement with Eqs.~\ref{eq_grado_medio_rectangulo_PB} and \ref{connection_area_rectangulo_PB}. We can also see that for a given value of $s>0$ and $N$, the largest expected average degree is obtained for PB-RGGs. Due to their periodic boundaries, PB-RGGs also present the largest (average) connection areas, when $s>0$ and $r=\sqrt{\pi}/(2\pi)$. Note that, according to Eqs.~\ref{eq_grado_medio} and \ref{eq_grado_medio_rectangulo_PB}, the expected (normalized) average degrees shown in Fig.~\subref*{comparacion_grados_medios_PERIODIC} are equal to the (average) connection areas of the respective graph models. On the other hand, we can see that RSGs exhibit the smallest expected average degree when $0 < s \lesssim 0.3$ and $a=2$, while for PB-RRGs this happens when $s \gtrsim 0.3$. According to Eq.~\ref{connection_area_rectangulo_PB}, as we elongate the rectangle, the range of $s$ values (i.e., $b/2 \leq s \leq a/2$) where the average expected degree of PB-RRGs and their connection area grow more slowly, is increased. Indeed, in the case of $a \gtrsim 1.95$, for a given value of $N$, there is a threshold connection radius, denoted as $s^T$, that makes the expected average degrees of RSGs and PB-RRGs equal. For instance,  when $a = 2$, $s^T=0.30814$ (see Fig.~\ref{grado_periodico}). Thus, when $s<s^T$ ($s>s^T$), the smallest expected average degree is obtained for RSGs (PB-RRGs). In Fig.~\subref*{otra_nueva_fig}, we show the dependence of $s^T$ on $a$. To do so, we solve numerically the following equation

\begin{eqnarray}
f^{PB}(s^T,a)=A_{SC_i}(s^T)=\left\{\begin{matrix}
0\leq s^T\leq \pi r & 2\pi r^2\left ( 1-\cos\left ( \frac{s^T}{r} \right ) \right ) \\
\pi r < s ^T & 4\pi r^2
\end{matrix}\right.,
\label{cal_sT}
\end{eqnarray}

\noindent for a spherical radius $r=\sqrt{\pi}/(2\pi)$. As can be observed, when $a\gtrsim 2$, $s^T \propto 1/a$.

The previous discussion on the expected average degrees of PB-RGGs and RSGs suggests that the other topological properties of these models will follow an opposite tendency when compared with the characteristics of RGGs and RSGs.  However, in the case of PB-RRGs and RSGs, the results indicate that the relative features of these models will depend on the value of $a$ (the length of the largest side of the PB-RRGs), which determines the value of $s^T$. When $s < s^T$, PB-RRGs are similar to PB-RGGs, whereas, when $s > s^T$, the PB-RRGs behave approximately like RRGs with a larger number of links. Hereafter, we will consider only the RGG and RRG models in our comparative study. An exhaustive research on the degree distribution, connectivity, average path length, diameter and clustering coefficient for PB-RGGs and PB-RRGs is being conducted and it will be published elsewhere.

\subsection{Connectivity}
\label{Connectivity}

In Graph Theory, a (simple) graph is connected when there is a path between every pair of vertices. In this section, we are interested in the RSG's probability of being connected as a function of the connection radius $s$. Our strategy is similar to the one used in Refs.~\cite{Estrada15a,Estrada17a}.

For the RGG, Penrose \cite{Penrose1997} proved that if
$M_N$ is the maximum length of an edge in the graph, then the
probability that $N \pi M_N^2 - \ln N \leq \alpha$ for a given $\alpha \in\mathbb{R}$ is

\begin{eqnarray}
\lim_{N\rightarrow \infty}P\left ( N\pi M_N^2-\ln N \leq \alpha \right )=\exp\left ( -\exp\left [ -\alpha \right ] \right ).
\label{def_Pen}
\end{eqnarray}

\noindent This means that $\alpha=\lim_{N\rightarrow \infty} -\ln(-\ln(P))$. Consequently, if the RGG is almost surely
connected (i.e. $P \rightarrow 1$), then $\alpha \rightarrow +\infty$,  when $N \rightarrow +\infty$. On the other hand, if the RGG is almost surely disconnected (i.e. $P \rightarrow 0$ ), then $\alpha \rightarrow -\infty$.

In the two dimensional case (i.e. $d=2$), when no boundaries effects are considered, it is possible to rewrite Eq.~\ref{def_Pen} as follows \cite{Estrada15a,Estrada17a}:


\begin{eqnarray}
\lim_{N\rightarrow \infty}P\left ( \bar{k}-\ln N \leq \alpha \right )=\exp\left ( -\exp\left [ -\alpha \right ] \right ),
\label{approx_pen}
\end{eqnarray}

\noindent where $\bar{k}$ is just the average degree of the RGG, given by Eq.~\ref{eq_grado_medio_rectangulo}.

In the previous section we have obtained an analytical expression for $\bar{k}$ in the RSG (Eq.~\ref{eq_grado_medio}). Taking adavantage of the fact that the sphere has no borders (i.e. there are no boundaries effects), we can replace its expected average degree in Eq.~\ref{approx_pen} and obtain an analogous expression for the RSGs. In this way, the RSGs' probability of being connected can be estimated as: 

\begin{eqnarray}
\lim_{N\rightarrow \infty}P\left ( \frac{N-1}{2}\left ( 1-\cos\left ( \frac{s}{r} \right ) \right )-\ln N \leq \alpha \right )=\exp\left ( -\exp\left [ -\alpha \right ] \right ).
\label{approx_pen_RSG}
\end{eqnarray}

According to Eq.~\ref{approx_pen_RSG}, a lower bound for the prior probability is given by

\begin{eqnarray}
\exp\left ( -\exp\left [ -\frac{N-1}{2}\left ( 1-\cos\left ( \frac{s}{r} \right ) \right ) +\ln N\right ] \right )  \leq \exp\left ( -\exp\left [ -\alpha \right ] \right ).
\label{connect_lim_esf}
\end{eqnarray}

\begin{figure}[h!]
\centering
\subfloat[]{\label{connectivity_500}\includegraphics[width=0.5\textwidth]{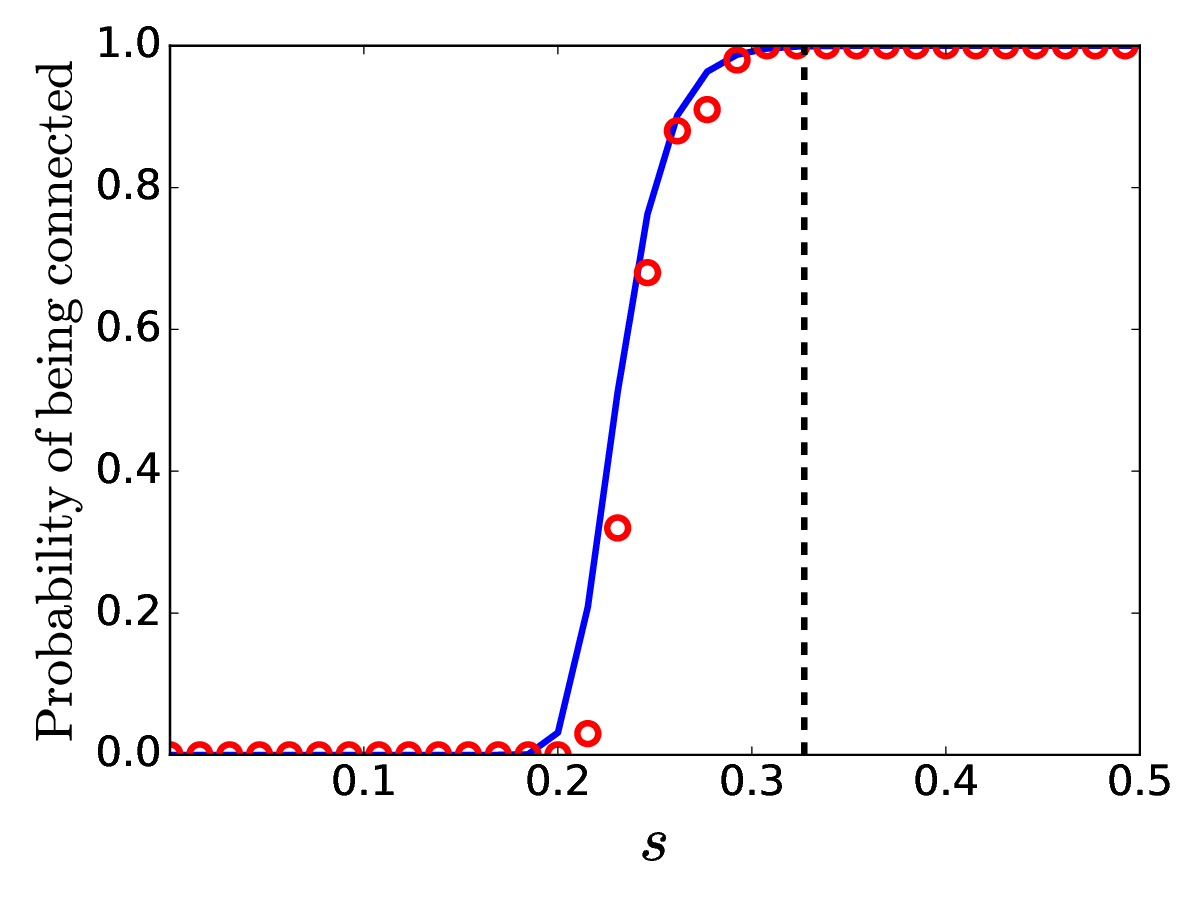}}
\subfloat[]{\label{connectivity_1000}\includegraphics[width=0.5\textwidth]{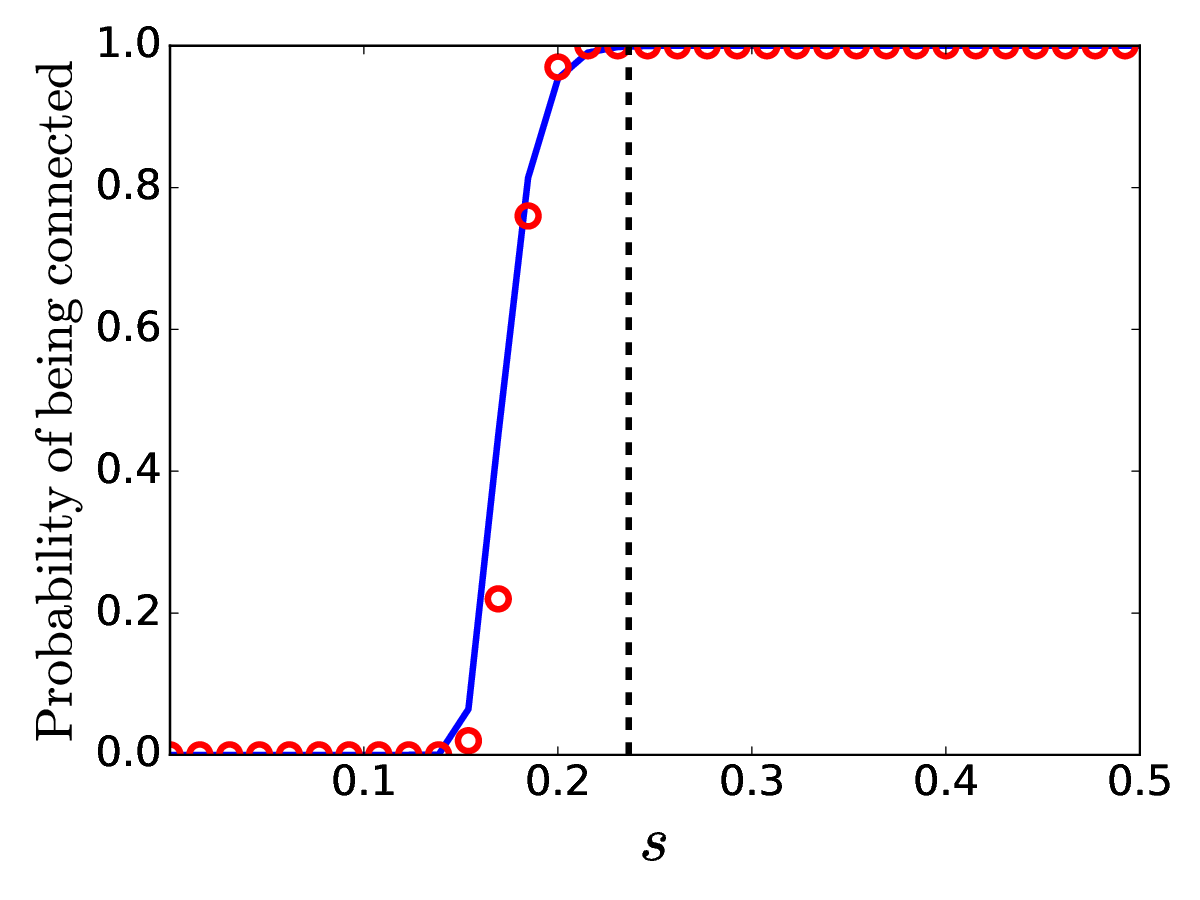}}
\caption{Dependence of the probability of being connected on $s$ for RSGs. The symbols indicate the numerical results (averaged over 100 realizations), the blue continuous line represents the results of Eq.~\ref{connect_lim_esf}, and the black dashed line shows the critical connection radius given by Eq.~\ref{s_crit}, when $\alpha_c=7$. (a) $N=500$. (b) $N=1000$.}
\label{fig_connectivity}
\end{figure}

In Fig.~\ref{fig_connectivity}, we compare the lower bound obtained from Eq.~\ref{connect_lim_esf} with the computer results (averaged over 100 realizations) for the dependence of the RSG' connectivity as a function of the connection radius $s$, when $r=1$. As can be seen, the observed points and the lower bound prediction follow the same distribution, and their respective values are very similar. As expected, the analytical and numerical results also show that, for a given connection radius $s$, the larger the number of nodes (i.e. the average degree), the larger the connectivity.

On the other hand, Eq.~\ref{approx_pen_RSG} allows us to approximate the connection radius at which the RSG is connected with probability one. Following the terminology in Ref.~\cite{Estrada17a}, we denote that value as \textit{critical connection radius}, $s_c$. In the case of RSGs, its lower bound can be estimated as:

\begin{eqnarray}
s_c^\mathrm{RSG} \geq r\arccos\left ( 1-\frac{2}{N-1}\left (\ln N +\alpha_c \right ) \right ),
\label{s_crit}
\end{eqnarray}

\noindent where $\alpha_c$ is the value of $\alpha$ for which $\exp\left ( -\exp\left [ -\alpha \right ] \right ) \approx 1$.
For instance, for $\alpha_c=7$ we obtain $\exp\left ( -\exp\left [ -\alpha \right ] \right ) \approx 0.999$.
Introducing this value of $\alpha_c$ into Eq.~\ref{s_crit} we obtain that $s_c^\mathrm{RSG} \gtrsim 0.327$ for RSGs with $N = 500$ nodes and $s_c^\mathrm{RSG} \gtrsim 0.237$ for those with $N = 1000$ nodes, when $r=1$ (unit ball). As can be seen in Fig.~\ref{fig_connectivity}, the previous values of $s_c^\mathrm{RSG}$ are in good agreement with our numerical findings.

Now we compare the probability of being connected and the critical connection radius obtained for RSGs with those for RGGs and RRGs. In the case of RRGs, using Eqs.~\ref{eq_grado_medio_rectangulo} and \ref{approx_pen}, we can derive the following lower bound for their probability of being connected:

\begin{eqnarray}
\exp\left ( -\exp\left [ -\frac{N-1}{(ab)^2}f(s,a) +\ln N\right ] \right )  \leq \exp\left ( -\exp\left [ -\alpha \right ] \right ).
\label{connect_lim_rect}
\end{eqnarray}

\begin{figure}[h!]
\centering
\subfloat[]{\label{connectivity_comp}\includegraphics[width=0.5\textwidth]{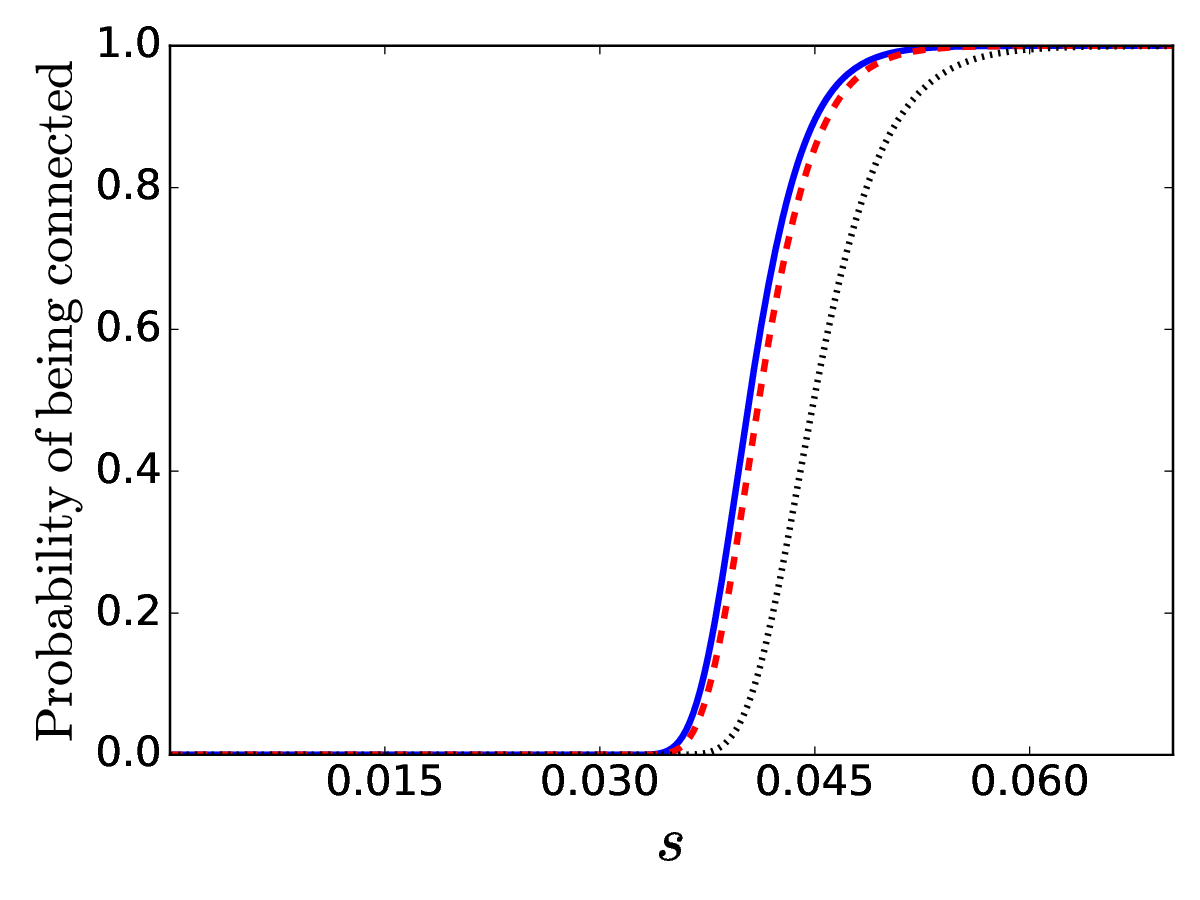}}
\subfloat[]{\label{radius_crit}\includegraphics[width=0.5\textwidth]{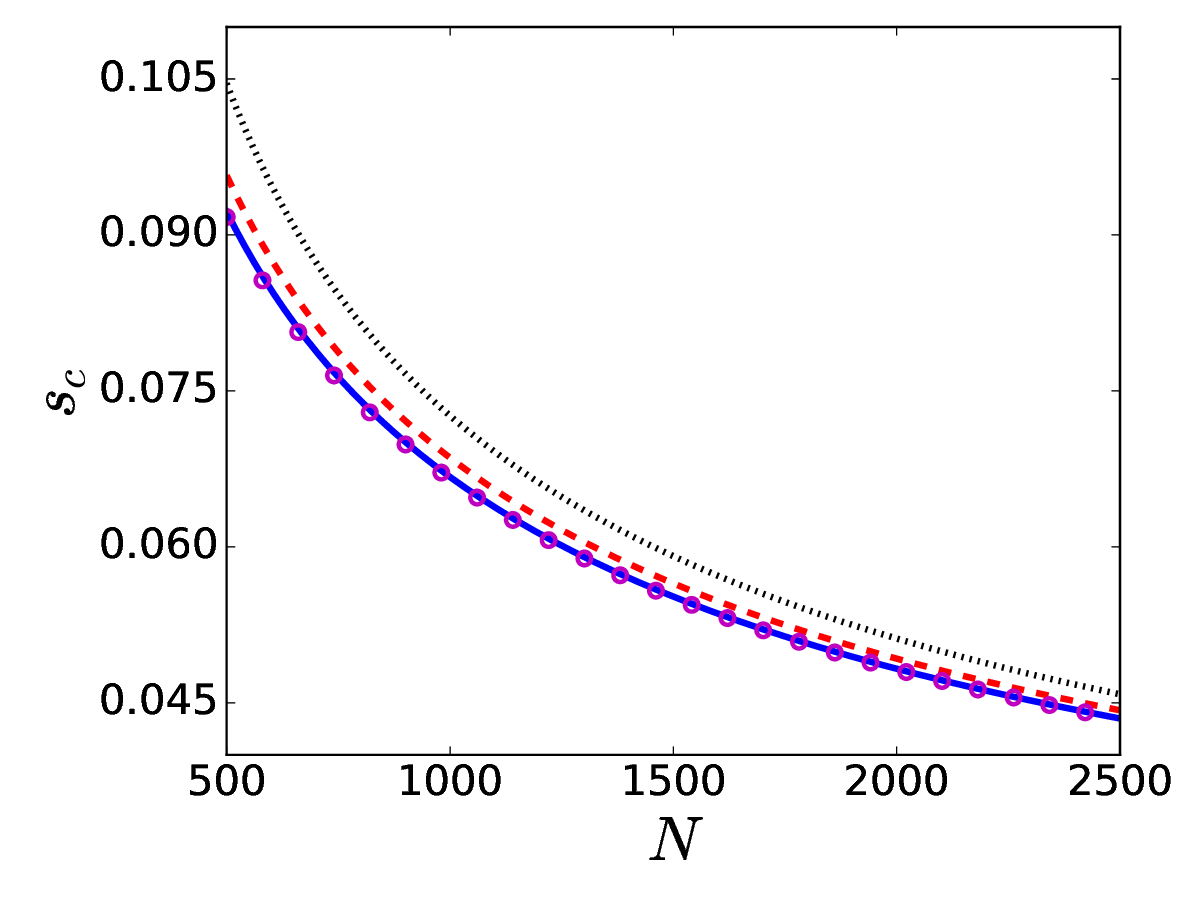}}
\caption{(a) Dependence of the probability of being connected on $s$ for RSGs with radius $r=\sqrt{\pi}/(2\pi)$ (blue continuous line, Eq.~\ref{connect_lim_esf}), RGGs (red dashed line, Eq.~\ref{connect_lim_rect}) and RRGs with $a=10$ and $b=0.1$ (black dotted line, Eq.~\ref{connect_lim_rect}), when $N=1500$. (b) Dependence of the critical connection radius $s_c$ on $N$ for a RSG (blue continuous line obtained from Eq.~\ref{s_crit}), RRG (red dashed line obtained from Eq.~\ref{s_crit_RRG2}, with $a=1$), RRG (black dotted line obtained from Eq.~\ref{s_crit_RRG2}, with $a=10$) and a RRG when no boundary effects are considered (magenta circles obtained from Eq.~\ref{s_crit_RRG}). In all previous cases, the value of $\alpha_c$ is equal to 7.}
\label{fig_connectivity_comp}
\end{figure}

In Fig.~\subref*{connectivity_comp}, we compare the analytical lower bounds for the connectivity of RSGs (with spherical radius $r=\sqrt{\pi}/(2\pi)$), RGGs and RRGs, respectively. As expected, for a given value of $s$ and a fixed value of $N$, the larger the length of the perimeter (i.e. the smaller the average degree), the smaller the connectivity. Therefore, the connectivity of RSGs (RRGs) is the largest (smallest) one.

On the other hand, considering that $\bar{k} = N\pi s^2$ when the boundary effects of the RRG are negligible (see Refs.~\cite{Estrada15a,Estrada17a}), the critical connection radius of the RRGs, $s_c^\mathrm{RRG}$, is bounded as follows:

\begin{eqnarray}
s_c^\mathrm{RRG}\leq \sqrt{\frac{\alpha _c+\ln N}{N\pi}}.
\label{s_crit_RRG}
\end{eqnarray}

\noindent If we take into account the boundary effects (i.e. $\bar{k} = (N-1)f(s,a)$), $s_c^\mathrm{RRG}$ is obtained from the following inequality:

\begin{eqnarray}
f(s_c^\mathrm{RRG},a)\leq \frac{\alpha _c+\ln N}{N-1}.
\label{s_crit_RRG2}
\end{eqnarray}

In Fig.~\subref*{radius_crit}, we show the dependence of the critical connection radius $s_c$ on $N$ for RSGs, RGGs and RRGs. As can be observed, for a given number of nodes and a fixed value of $\alpha_c$, the sphere (the rectangle) exhibits the smallest (largest) value of $s_c$, when the boundary effects are considered (see Eqs.~\ref{s_crit} and \ref{s_crit_RRG2}). It is also possible to see that the larger the number of nodes, the smaller the differences between the systems. On the other hand, in case we ignore the boundary effects of the RRGs, as well as those of RGGs (Eq.~\ref{s_crit_RRG}), we found no significant differences between the critical connection radii of the various systems considered, when $N$ is large enough ($N>500$).

\subsection{Degree distribution}
\label{deg_distribution}

To estimate the degree distribution of RSGs we follow a strategy that is similar to the one used in Refs.~\cite{Estrada15a,Estrada17a,Antonioni12}. We consider the probability of having a node $i$ with degree $k_i$ when there are $N-1$ other nodes uniformly distributed in the sphere. When N is large enough, $N-1 \sim N$ and this probability follows the binomial distribution:

\begin{eqnarray}
P(k)=\binom{n}{k}p^k(1-p)^{n-k},
\end{eqnarray}

\noindent where $p$ is the connection area of the node $i$, given by Eq.~\ref{area_cap}. When $N\rightarrow \infty$ and $s$ is small enough, $P(k)$ approaches a Poisson distribution of the form:

\begin{eqnarray}
P(k)\approx\frac{\bar{k}^k e^{-\bar{k}}}{k!}.
\label{eq_poisson}
\end{eqnarray}

\begin{figure}[h!]
\centering
\includegraphics[width=0.5\textwidth]{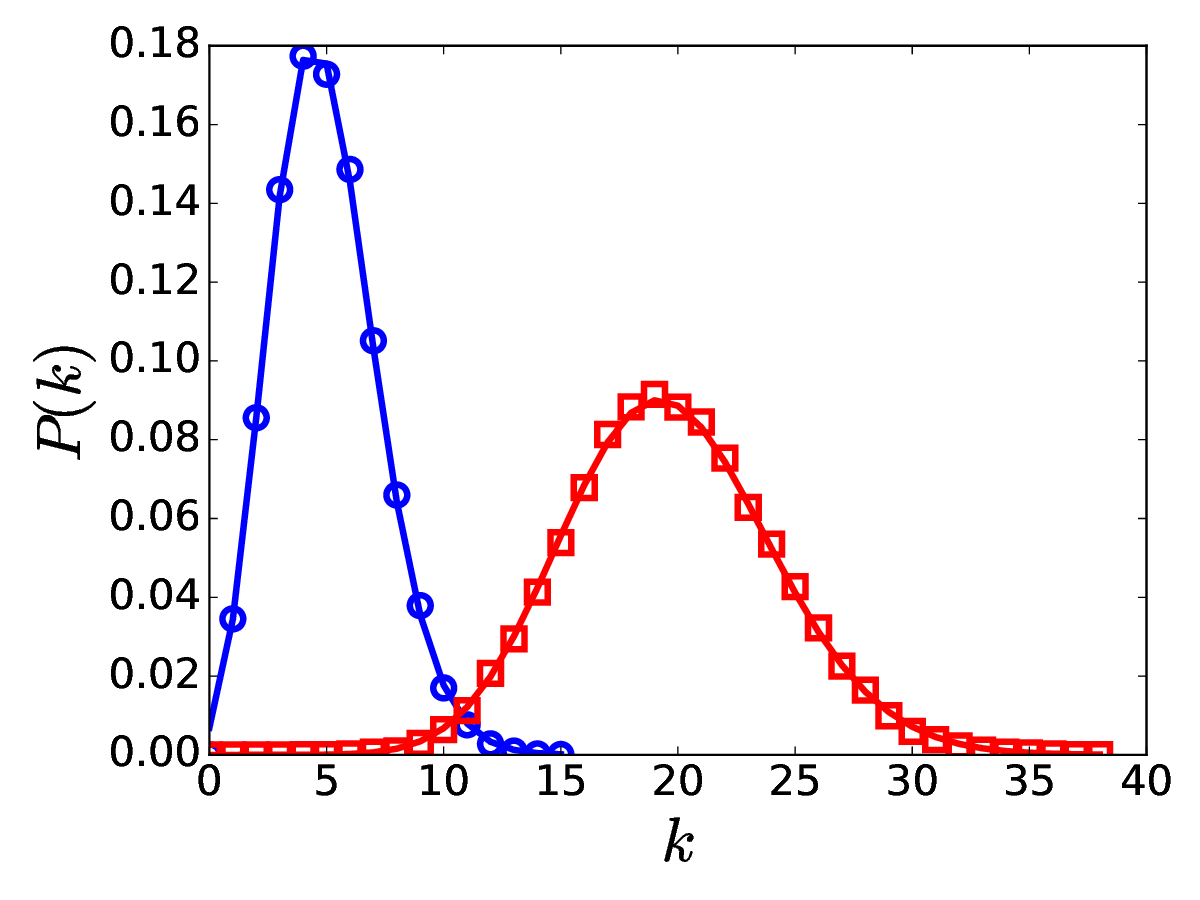}
\caption{Degree distribution $P(k)$ of RSGs with $N=500$ nodes and connection radius $s = 0.2$ (blue circles) and $s=0.4$ (red squares). The symbols indicate the numerical results (averaged over 100 realizations). The continuous lines correspond to the shape of the Poisson distribution (Eq.~\ref{eq_poisson}) with the
corresponding average degree obtained from Eq.~\ref{eq_grado_medio}.}
\label{degree_distribution}
\end{figure}

\noindent Thus, introducing the analytical expression for the average degree of the RSGs (Eq.~\ref{eq_grado_medio}) into Eq.~\ref{eq_poisson}, we can easily approximate the degree distribution of these graphs. In Fig.~\ref{degree_distribution} we show the results for $P(k)$, when $N=500$ and two connection radii are used. As can be observed, the numerical findings (averaged over 100 random realizations) are in excellent agreement with the results obtained from Eq.~\ref{eq_poisson}. Therefore, it is possible to see that spherical geometry of RSGs does not affect the shape of the degree distribution estimated in Refs.~\cite{Estrada15a,Estrada17a,Antonioni12}.

\begin{figure}[h!]
\centering
\includegraphics[width=0.5\textwidth]{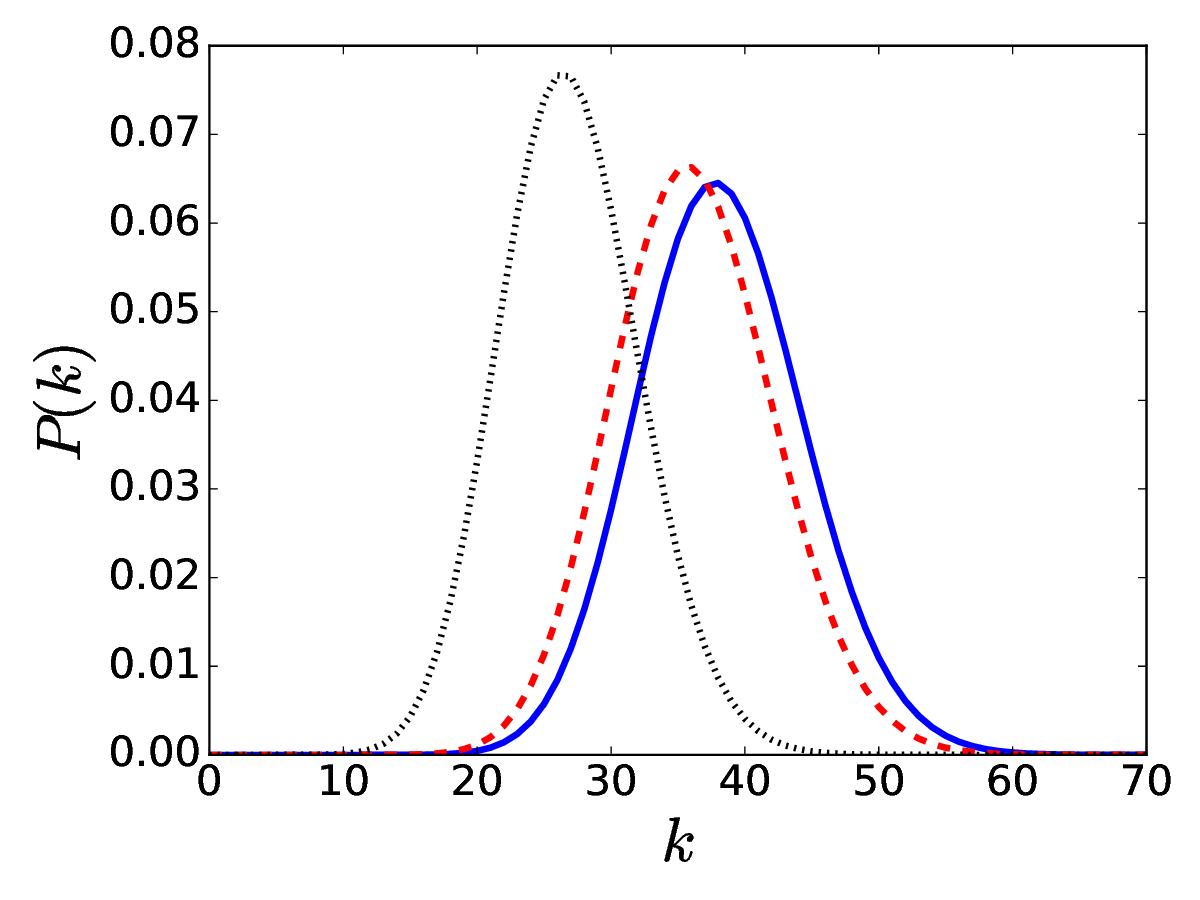}
\caption{Analytical degree distribution $P(k)$ for RSGs with $r=\sqrt{\pi}/(2\pi)$ (blue continuous line), RGGs with $a=1$ and $b=1$ (red dashed line), and RRGs with $a=10$ and $b=0.1$ (black dotted line), when $N=2500$ nodes and $s=0.07$. The lines correspond to the shape of the Poisson distribution (Eq.~\ref{eq_poisson}) with the corresponding average degrees, obtained from Eq.~\ref{eq_grado_medio} (RSGs) and Eq.~\ref{eq_grado_medio_rectangulo} (RGGs and RRGs), respectively.}
\label{degree_distribution_comp}
\end{figure}

Now we compare the degree distribution obtained for the RSGs with those for RGGs and RRGs. To calculate the latters, we introduce the average degree of the RRGs (Eq.~\ref{eq_grado_medio_rectangulo}) into Eq.~\ref{eq_poisson}. In Fig.~\ref{degree_distribution_comp} we exhibit the analytical degree distribibution for RSGs (with spherical radius $r=\sqrt{\pi}/(2\pi)$), RGGs and RRGs, when $N=2500$ nodes and $s=0.07$. We can observe that the results are in perfect agreement with the findings we presented in the previous sections. That is, for a given connection radius $s$ and a fixed number of nodes $N$, the larger the length of the perimeter of the surface, the smaller the average degree of the corresponding network.

\subsection{Average shortest path distance}
\label{shortest_path}

Let $G(V,E)$ be a simple connected graph. A \textit{path} of length $k$ in $G$ is a set of nodes $i_1$, $i_2$, $\cdots$, $i_k$, $i_{k+1}$ such that for all $1\leq \ell \leq k$, $(i_\ell,i_{\ell+1}) \in E$ with no repeated nodes. The \textit{shortest-path} between two nodes $i,j \in V$ is defined as the length of the shortest path connecting these nodes. We denote this shortest path distance as $d_G(i,j)$. Then, the \textit{average shortest path length} is defined as follows:

\begin{eqnarray}
\left \langle \ell \right \rangle = \frac{1}{N(N-1)} \sum_{i,j\in V} d_G(i,j).
\end{eqnarray}

\noindent In the case that $G(V,E)$ is disconnected, $\left \langle \ell \right \rangle=\infty$.

\begin{figure}[h!]
\centering
\includegraphics[width=0.35\textwidth]{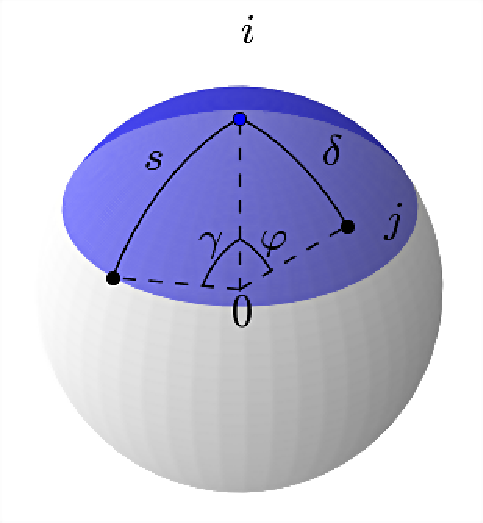}
\caption{Example of two points $i,j \in S^2$ that are separated by a great circle distance $\delta(i,j)=r\varphi$. In this case, $\delta(i,j)<s$ and, consequently, $d_G^{GC}(i,j)=1$ (Eq.~\ref{shrt_path_GC}).}
\label{diag_shortest}
\end{figure}

We are interested in determining a bound for the RSGs' average shortest path length as a function of the connection radius $s$. Let $i$ and $j$ be two points in the sphere separated by a great circle distance $\delta(i,j)=r\varphi$ and let $s$ be the connection radius of the RSG (see Fig.~\subref*{cap}). Let us now define the \textit{great-circle shortest-path} between two nodes $i,j \in V$ as follows:

\begin{eqnarray}
d_G^{GC}(i,j)=\left \lceil \frac{r\varphi}{s} \right \rceil,
\label{shrt_path_GC}
\end{eqnarray}

\noindent where $\left \lceil . \right \rceil$ represents the ceiling function. Supposing that $N\rightarrow \infty$, there will be a large amount of nodes in the great circle arc that connects $i$ with $j$. Therefore, it is almost certain that, in the case of $s\ll \pi r$, $d_G(i,j)=d_G^{GC}(i,j)$. Now, suppose that node $i$ is located in the North pole of the sphere (see Fig.~\ref{diag_shortest}). Then, for a given value of $s$, we can estimate the average great-circle shortest-path between $i$ and all the other points in the surface of the sphere, as follows

\begin{eqnarray}
\left \langle \ell ^\mathbf{GC}\right \rangle(s)=
\frac{\int_{\theta=0 }^{\theta=2\pi }\int_{\varphi=0 }^{\varphi=\pi} d_G^{GC} r^2\sin(\varphi'){\mathrm{d} \varphi'}{\mathrm{d} \theta'}}{\int_{\theta=0 }^{\theta=2\pi }\int_{\varphi=0 }^{\varphi=\pi} r^2\sin(\varphi'){\mathrm{d} \varphi'}{\mathrm{d} \theta'}}
=\frac{\int_{\theta=0 }^{\theta=2\pi }\int_{\varphi=0 }^{\varphi=\pi} \left \lceil \frac{r\varphi'}{s} \right \rceil r^2\sin(\varphi'){\mathrm{d} \varphi'}{\mathrm{d} \theta'}}{4\pi r^2}.
\label{lim_inf_shortest}
\end{eqnarray}

\noindent In the case that $N$ is not large enough, there won't be enough nodes in the great circle arc that connects $i$ with $j$. Consequently, if the connection radius $s$ is small, then $d_G(i,j) \geq d_G^{GC}(i,j)$. Thus, $\left \langle \ell ^\mathbf{GC}\right \rangle$ is the lower bound for $\left \langle \ell \right \rangle$.

\begin{figure}[h!]
\centering
\subfloat[]{\label{shortest}\includegraphics[width=0.5\textwidth]{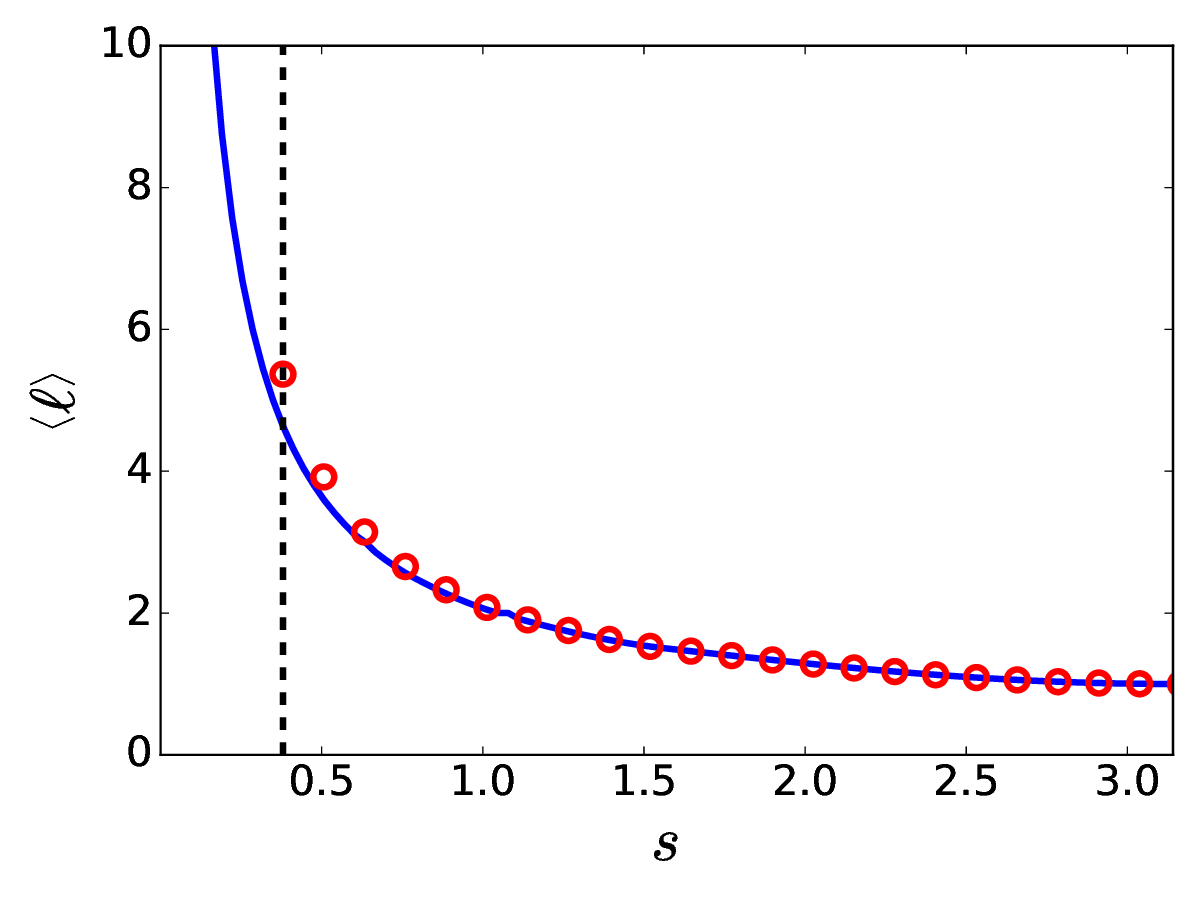}}
\subfloat[]{\label{shortest_comp}\includegraphics[width=0.5\textwidth]{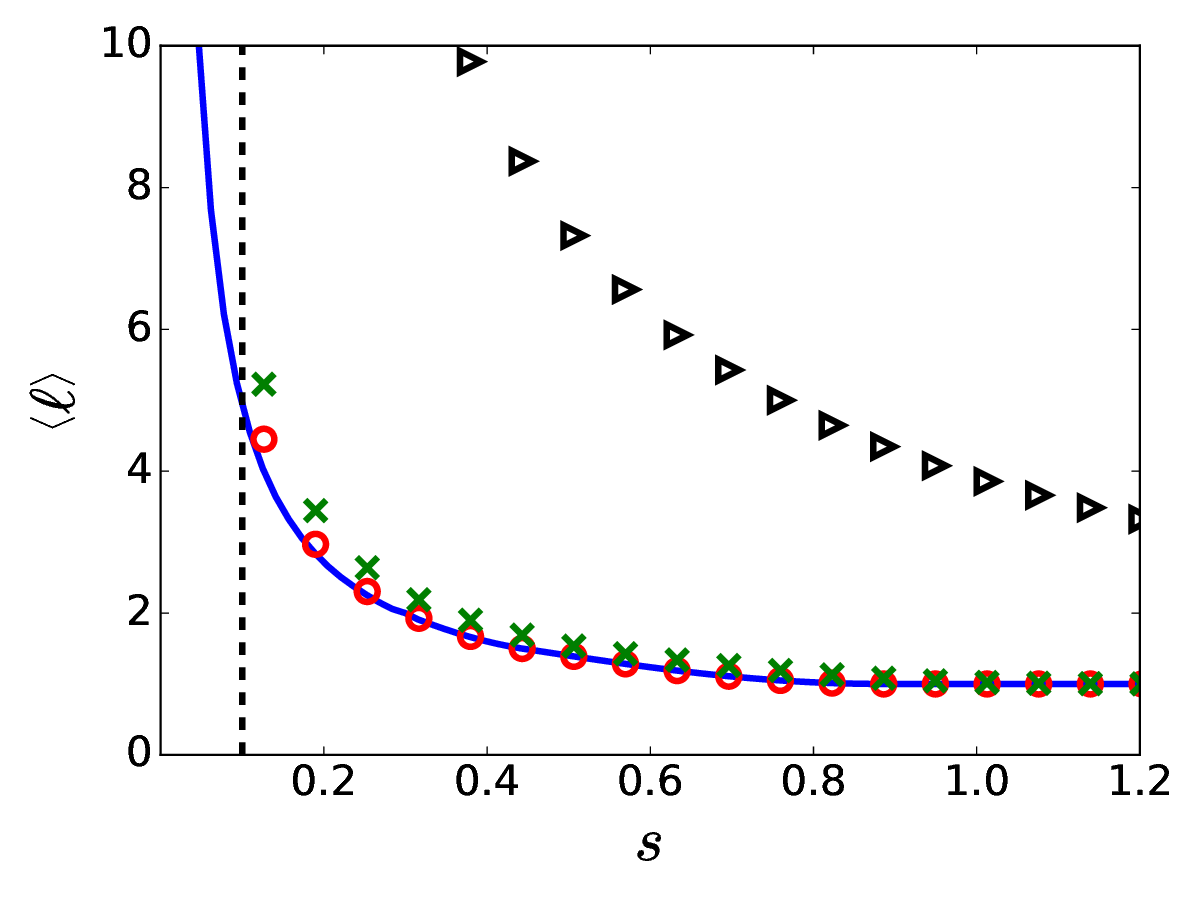}}
\caption{(a) Dependence of the average shortest path length $\left \langle \ell \right \rangle$ on $s$ for a RSG with $N=500$ nodes, when $r=1$. The symbols indicate the numerical results (averaged over 100 realizations). The blue continuous line represents the numerical evaluation of the lower bound for  $\left \langle \ell \right \rangle$ (Eq.~\ref{lim_inf_shortest}), and the black dashed line shows 
the critical connection radius $s_c^\mathrm{RSG}$ (Eq.~\ref{s_crit}), when $\alpha_c=7$.
(b) Dependence of the average shortest path length $\left \langle \ell \right \rangle$ on $s$ for a RSG with $r=\sqrt{\pi}/(2\pi)$ (red circles), RRG with $a=1$ (green crosses), and RRG with $a=10$ (black triangles), when $N=500$ nodes. The blue continuous line represents the numerical evaluation of the lower bound for the RSG's shortest path length (Eq.~\ref{lim_inf_shortest}), and the black dashed line shows 
the critical connection radii for the three networks considered, $s_c \approx 0.1$ (see Fig.~\ref{fig_connectivity_comp}(b)), when $\alpha_c=7$.}
\label{fig_shortest}
\end{figure}

In Fig.~\subref*{shortest} we compare the numerical results for the average shortest path $\left \langle \ell \right \rangle$ with our lower bound (Eq.~\ref{lim_inf_shortest}), when $r=1$. As we can see, when $s \geq s_c^\mathrm{RSG}$, the lower bound is very close to the average shortest path obtained for these RSGs. Particularly, for large values of the connection radius, the computer results are almost identical to those of the analytical lower bound. This behavior takes place because the larger the value of $s$, the smaller the amount of nodes in the great circle arc that are needed to connect two nodes (i.e. rising $s$ increases the similarity between $d_G(i,j)$ and $d_G^{GC}(i,j)$). On the other hand, as we discussed in the previous sections, in case of $s<s_c^\mathrm{RSG}$, the network is almost surely disconnected, and therefore, $\left \langle \ell \right \rangle=\infty$.


When we compare the  average shortest path length obtained for the RSGs with those for RGGs and RRGs (see Fig.~\subref*{shortest_comp}), we can observe that the evolution of $\left \langle \ell \right \rangle$ for RGGs and RRGs exhibits a monotonically decreasing behavior similar to that of RSGs. The sphere with $r=\sqrt{\pi}/(2\pi)$ (the rectangle with $a>1$) shows the smallest (largest) average shortest path length for $s \geq s_c$. For a given value of $s$ and a fixed $N$, the larger the length of the perimeter, the larger the average shortest path length. For a wider description regarding the influence of $a$ on the shortest path length of RRGs the reader is referred to Ref. \cite{Estrada15a}.

Using similar arguments, we can estimate the \textit{diameter} of the RSGs as a function of the connection radius $s$. Given a simple connected graph $G(V,E)$, the diameter is defined as the maximum of all the shortest path lengths in the graph, i.e. $D = \max_{i,j} d_G(i,j)$. Supposing that $N\rightarrow \infty$, the diameter represents the shortest path length between two nodes that are separated by a great circle distance $\delta(i,j)=\pi r$ (see Fig.~\ref{diag_shortest}). Therefore, it can be approximated by

\begin{eqnarray}
D^{GC}=\left \lceil \frac{\pi r}{s} \right \rceil.
\label{diameter_GC}
\end{eqnarray}

\noindent If $N$ and $s$ are not large enough, then $D^{GC}\leq D$ (i.e. $D^{GC}$ is a lower bound for the diameter of RSGs).

\begin{figure}[h!]
\centering
\subfloat[]{\label{diam_N_500_uniball}\includegraphics[width=0.5\textwidth]{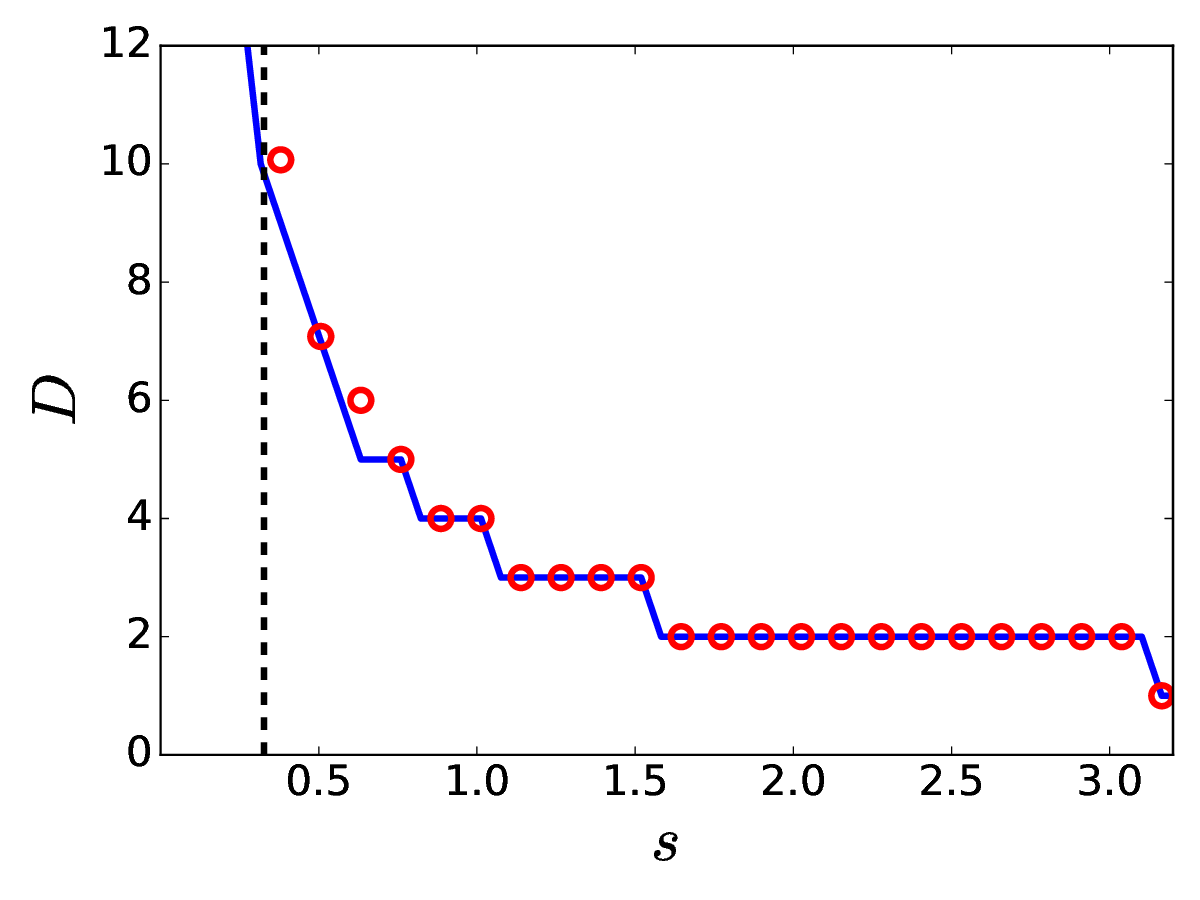}}
\subfloat[]{\label{diameter_comp}\includegraphics[width=0.5\textwidth]{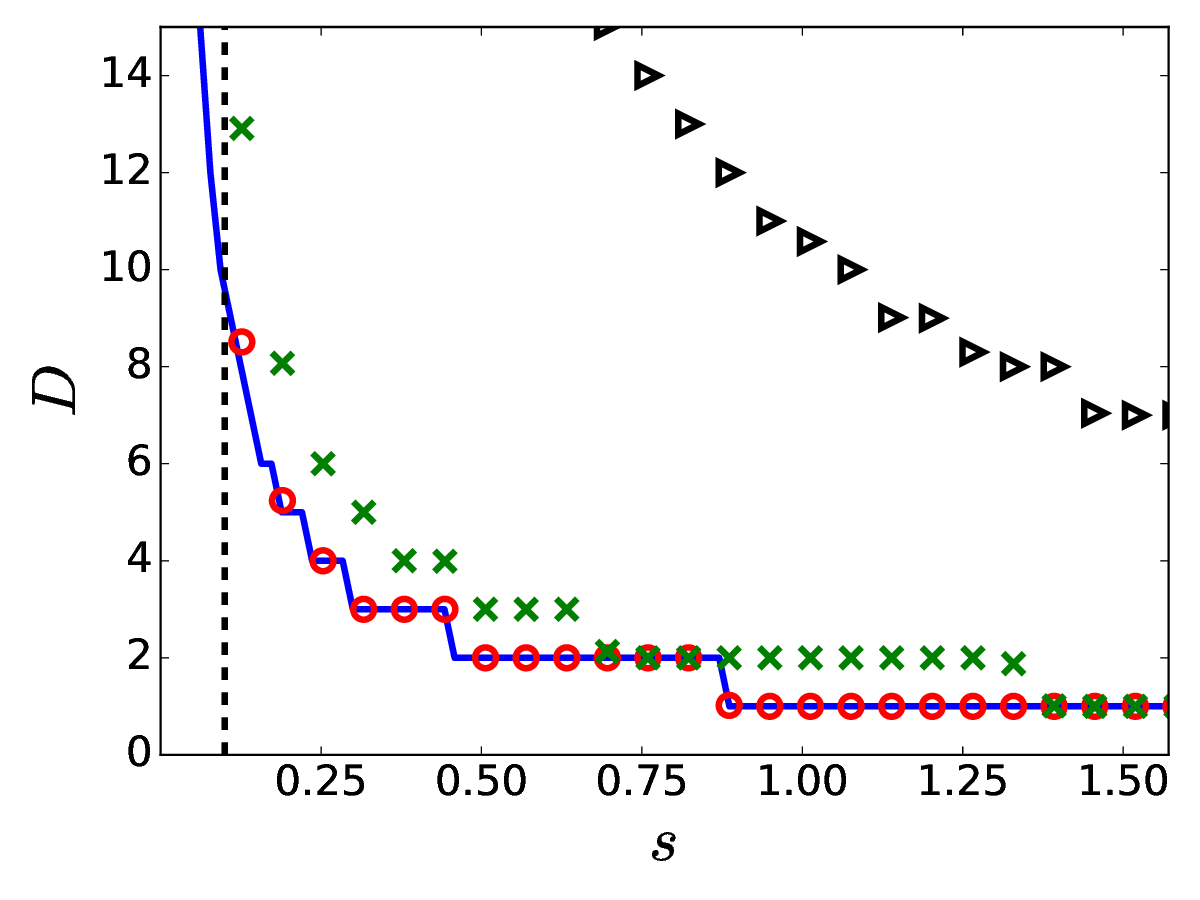}}
\caption{(a) Dependence of the diameter $D$ on $s$ for a RSG with $N=500$ nodes, when $r=1$. The symbols indicate the numerical results (averaged over 100 realizations). The blue continuous line represents the numerical evaluation of the lower bound for  $\left \langle \ell \right \rangle$ (Eq.~\ref{diameter_GC}), and the black dashed line shows 
the critical connection radius $s_c^\mathrm{RSG}$ (Eq.~\ref{s_crit}), when $\alpha_c=7$.
(b) Dependence of the average diameter $D$ on $s$ for a RSG with $r=\sqrt{\pi}/(2\pi)$ (red circles), RRG with $a=1$ (green crosses), and RRG with $a=10$ (black triangles), when $N=500$ nodes. The blue continuous line represents the numerical evaluation of the lower bound for the RSG's shortest path length (Eq.~\ref{diameter_GC}), and the black dashed line shows 
the critical connection radii for the three networks considered, $s_c \approx 0.1$ (see Fig.~\ref{fig_connectivity_comp}(b)), when $\alpha_c=7$.}
\label{fig_diameter}
\end{figure}

In Fig.~\subref*{diam_N_500_uniball} we compare the numerical results for the diamater of the RSG with our lower bound (Eq.~\ref{diameter_GC}), when $r=1$. When $s \geq s_c^\mathrm{RSG}$, the lower bound is very similar to the estimated diameter, and, for large connection radius, they are almost identical (for the same reason as the shortest path length). Thus, according to Eq.~\ref{diameter_GC}, given a sphere with radius $r$, the diameter of its RSGs depends only on the inverse of the connection radius, $s$. Note that, in case of $s<s_c^\mathrm{RSG}$, the network is almost surely disconnected, and consequently, $\left \langle D \right \rangle=\infty$.


Finally, in Fig.~\subref*{diameter_comp}, we compare the diameter obtained for the RSGs with those for RGGs and RRGs. As expected, the evolution of $D$ for RGGs and RRGs shows a monotonically decreasing behavior similar to that of RSGs. The sphere with $r=\sqrt{\pi}/(2\pi)$ (the rectangle with $a>1$) shows the smallest (largest) diameter for $s \geq s_c$. For a given value of $s$ and a fixed $N$, the larger the length of the perimeter, the larger the diameter.

\subsection{Clustering coefficient}
\label{clustering}

According to Ref.~\cite{Watts1998440}, the \textit{local clustering coefficient} of a node $i$ is given by

\begin{eqnarray}
C_i = \frac{2t_i}{k_i(k_i-1)}=\frac{\left ( \mathbf{A}^3 \right )_{ii}}{k_i(k_i-1)},
\label{clustering_coeff_def}
\end{eqnarray}

\noindent where $t_i$ is the number of triangles incident to the node $i$, $\mathbf{A}$ is the adjacency matrix of the network, and $k_i$ is the degree of the node $i$. Then, the \textit{average clustering coefficient} of the network is defined as

\begin{eqnarray}
\left \langle C \right \rangle =\frac{1}{N}\sum _{i=1}^N C_i,
\label{aver_clustering_coeff_def}
\end{eqnarray}

\noindent for $k_i > 1$. In the case of $k_i = 1$, the term $2t_i / k_i (k_i - 1)$ is considered equal to zero.

\begin{figure}[h!]
\centering
\includegraphics[width=0.35\textwidth]{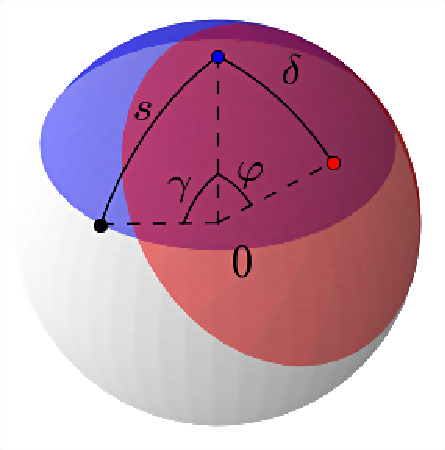}
\caption{Overlapping rea of two spherical caps with connection ratius $s$ that are separated by a distance $\delta$.}
\label{solapamiento}
\end{figure}

We are interested in determining a bound for the RSGs' average clustering coefficient as a function of the connection radius $s$, similar to the strategy used in Refs.~\cite{Dall02,Estrada15a}. Suppose $N \rightarrow \infty$, and let $i$ and
$j$ be two connected nodes in a RSG, which are separated at a
great circle distance $\delta(i,j)=\varphi r$ from each other. Let us draw two spherical caps
of connection radius $s$ centered, respectively, at $i$ and $j$. Let $\delta(i,j)\leq s$ such
that the two nodes are connected. Then, because $\delta(i,j)\leq 2s$ the
two caps overlap (see Fig.~\ref{solapamiento}). Because $i$ and $j$ are connected, any point
in the area formed by the overlap of the two caps will form
a triangle with the nodes $i$ and $j$. In addition, any node inside
the two caps that is not in the overlapping area forms a path
of length two with the nodes $i$ and $j$ . Thus, if we quantify
the ratio of the overlapping area and the total area of the cap,
we account for the ratio of the number of triangles and open
triads in which the nodes $i$ and $j$ take place, i.e., the clustering
coefficient. This ratio is given by

\begin{eqnarray}
C_i(\delta,s)=\frac{{A\bigcap_{SC_1}^{SC_2}}(\delta,s)}{2\pi r^2(1-\cos(s))}
\end{eqnarray}

\noindent where ${A\bigcap_{SC_1}^{SC_2}}(\delta,s)$ represents the overlapping area of two spherical caps with connection radius $s$ that are separated by a distance $\delta\leq 2s$, defined by

\begin{eqnarray}
{A\bigcap_{SC_1}^{SC_2}}(\delta,s) =
\left\{\begin{matrix}
0 = s & 0\\ 
0 < s \leq \pi/2 & 2\pi r^2 -2 r^2\arccos \left ( \cos(\delta)\csc^2 (s) -\cot^2(s)\right ) \\ 
& -4 r^2\arccos\left ( \csc(\delta)\cot(s) -\cot(\delta)\cot(s) \right )\cos(s) \\ 
\pi/2 < s \leq \pi, & 2\pi  r^2(1-2\cos(s))-2  r^2\arccos \left ( \cos(\delta)\csc^2 (\pi-s) -\cot^2(\pi-s)\right ) \\ 
& -4 r^2\arccos\left ( \csc(\delta)\cot(\pi-s) -\cot(\delta)\cot(\pi-s) \right )\cos(\pi-s)
\end{matrix}\right.
\label{eq_solap_caps}
\end{eqnarray}

\noindent Using the formulas in Ref.~\cite{Lee14}, Eq.~\ref{eq_solap_caps} can be derived.


Suppose node $i$ is located in the North pole of the sphere (see Fig.~\ref{solapamiento}). Now, in case of $\delta\leq2s$, for a given value of $s$, we estimate the average clustering coefficient between $i$ and all the other points in the cap $SC(i,s,r)$, as follows

\begin{eqnarray}
\left \langle C^{U} \right \rangle=\frac{\int_{\theta=0 }^{\theta=2\pi }\int_{\varphi=0 }^{\varphi=s} C_i(\varphi',s)r^2\sin(\varphi'){\mathrm{d} \varphi'}{\mathrm{d} \theta'}}{\int_{\theta=0 }^{\theta=2\pi }\int_{\varphi=0 }^{\varphi=s} r^2\sin(\varphi'){\mathrm{d} \varphi'}{\mathrm{d} \theta'}}=
\frac{\int_{\theta=0 }^{\theta=2\pi }\int_{\varphi=0 }^{\varphi=s} \left ( {A\bigcap_{SC_1}^{SC_2}}(\varphi',s) \right ) \sin(\varphi'){\mathrm{d} \varphi'}{\mathrm{d} \theta'}}{4\pi^2 r^2(1-\cos(s))^2}.\nonumber\\
\label{average_clustering_RSG}
\end{eqnarray}

\noindent If $N$ and $s$ are not large enough, the node's distribution on the surface of the sphere can not be considered homogeneous. Consequently, $\left \langle C^{U} \right \rangle \geq \left \langle C \right \rangle$ (i.e. $\left \langle C^{U} \right \rangle$ is an upper bound for the average clustering coefficient of RSGs).

\begin{figure}[h!]
\centering
\subfloat[]{\label{clustering_500}\includegraphics[width=0.5\textwidth]{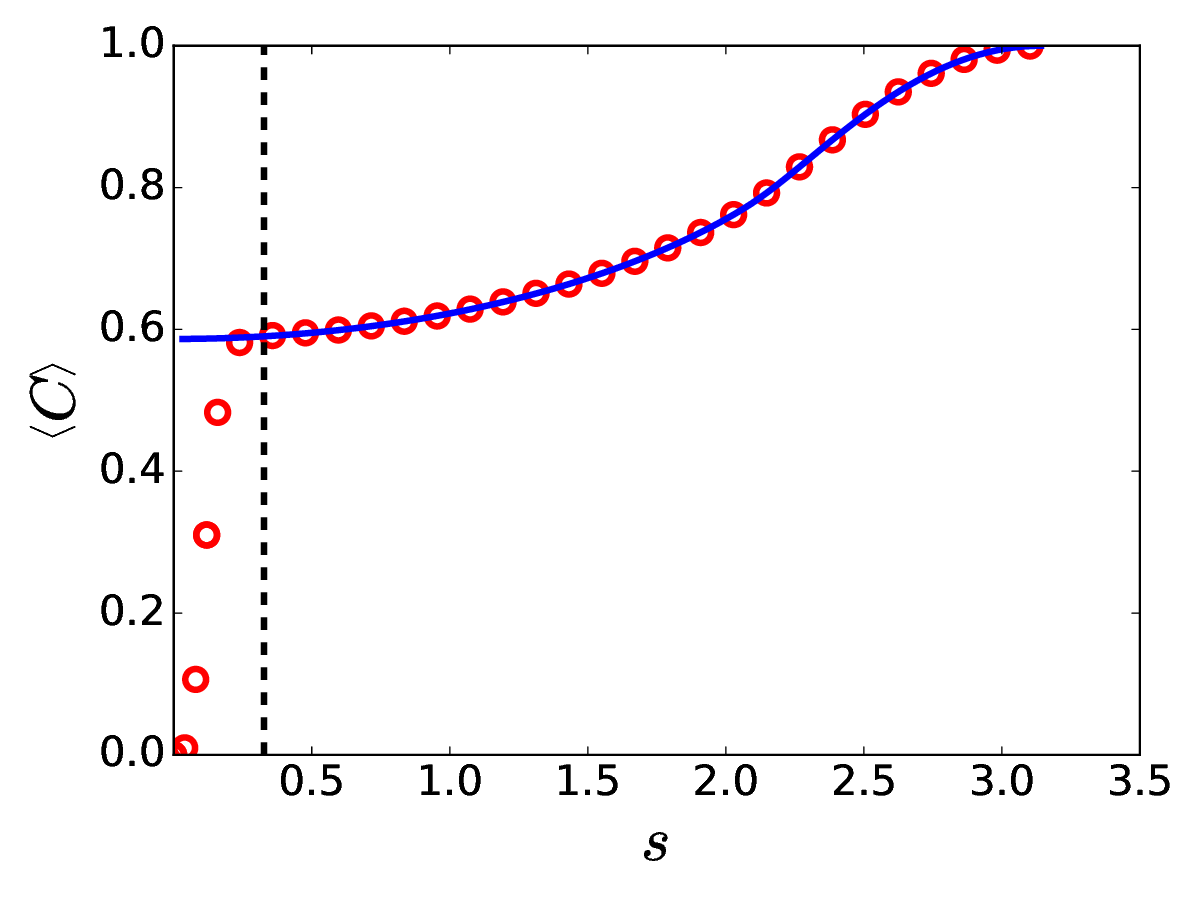}}
\subfloat[]{\label{clustering_comp}\includegraphics[width=0.5\textwidth]{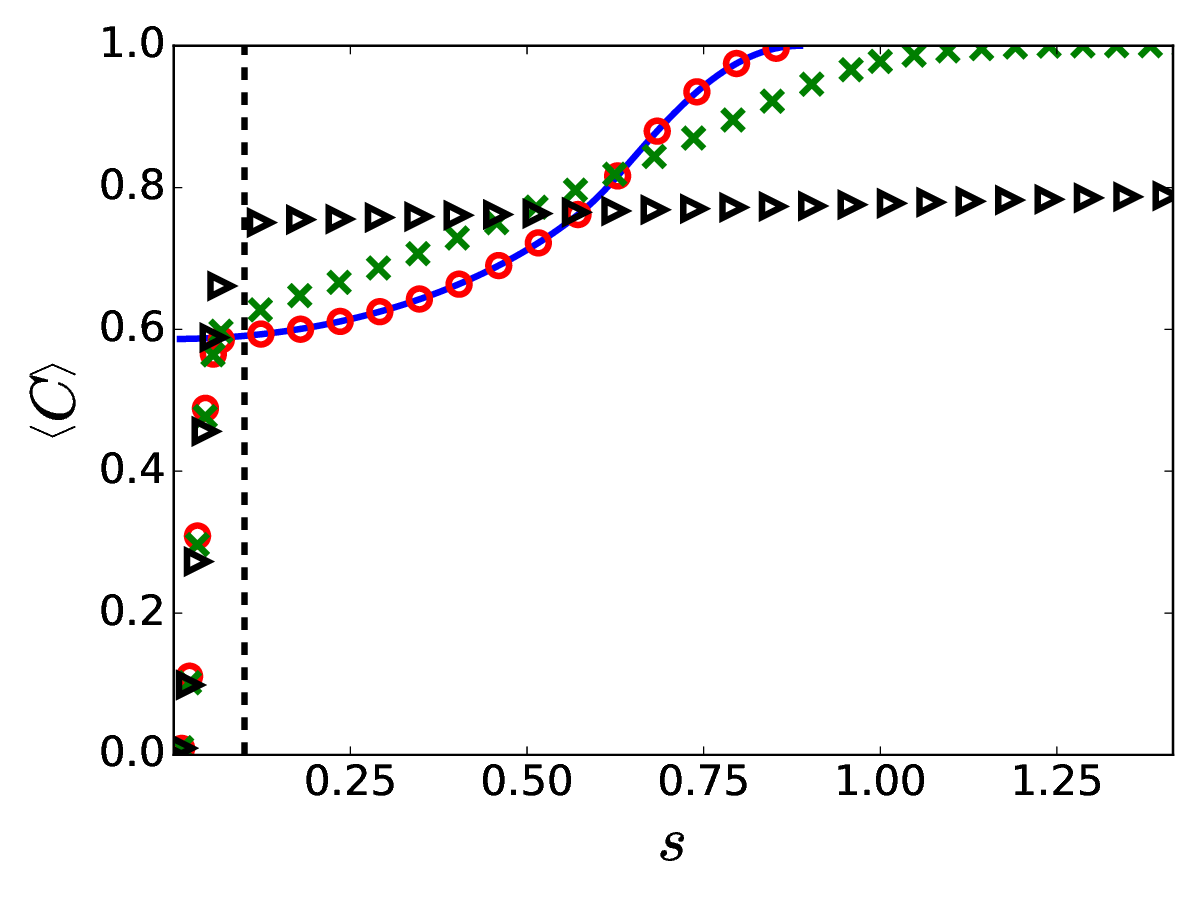}}
\caption{(a) Dependence of the average clustering coefficient $\left \langle C \right \rangle$ on the connection radius $s$ for RSGs with $N=500$ nodes and $r=1$. The symbols indicate the numerical results (averaged over 100 realizations) obtained from Eq.~\ref{aver_clustering_coeff_def}. The blue continuous line represents the numerical evaluation of the upper bound for $\left \langle C \right \rangle$ (Eq.~\ref{average_clustering_RSG}). The black dashed line represents 
the critical connection radius $s_c^\mathrm{RSG}$ (Eq.~\ref{s_crit}), when $\alpha_c=7$.
(b) Dependence of the average clustering coefficient $\left \langle C \right \rangle$ on $s$ for a RSG with $r=\sqrt{\pi}/(2\pi)$ (red circles), RRG with $a=1$ (green crosses), and RRG with $a=10$ (black triangles), when $N=500$ nodes. The blue continuous line represents the numerical evaluation of the upper bound for $\left \langle C \right \rangle$ (Eq.~\ref{average_clustering_RSG}), and the black dashed line shows 
the critical connection radii for the three networks considered, $s_c \approx 0.1$ (see Fig.~\ref{fig_connectivity_comp}(b)), when $\alpha_c=7$.}
\label{fig_clustering}
\end{figure}

In Fig.~\subref*{clustering_500} we exhibit the numerical results for the average clustering coefficient $\left \langle C \right \rangle$ and those for our lower bound (Eq.~\ref{average_clustering_RSG}), when $r=1$. As can be observed, when $s \gtrsim s_c^{\mathrm{RSG}}$, the estimated values of $\left \langle C \right \rangle$ are identical to those of the upper bound $\left \langle C^U \right \rangle$. On the other hand, it is remarkable that in the case of $s\rightarrow 0^+$, the numerical estimation of Eq.~\ref{average_clustering_RSG} converges to the analytical average clustering coefficient for two-dimensional RGGs obtained by Dall and Christensen in Ref.~\cite{Dall02}, when $N\rightarrow \infty$ and  $\alpha\rightarrow \infty$ (see Sec. \ref{Connectivity}). That is,

\begin{eqnarray}
\lim_{s\rightarrow 0^+} \left \langle C^U \right \rangle = 0.58653 \approx \frac{3\sqrt{3}}{4\pi}.
\end{eqnarray}

\noindent Intuition suggests that this behavior takes place because the surfaces of the small caps look flat when compared to the surface of the sphere. As expected, the data show that the increase of the connection radius $s$ augments the average clustering coefficient. Increasing $s$ enlarge the RSG's average degree and makes the network more dense. Indeed, the value of $\left \langle C \right \rangle$ exhibits an abrupt increase from $s=0$ (i.e. a graph without edges) to $s=s_c$ (i.e. a connected graph). For large values of $s$, the average clustering coefficient approachs to one. Finally, in case of $s \geq \pi r$, the RSG becomes a complete graph (i.e. a network in which every pair of distinct nodes is connected by a unique link).


In Fig.~\subref*{clustering_comp}, we compare the average clustering coefficient obtained for the RSGs with those for RGGs and RRGs. As expected, the evolution of $\left \langle C \right \rangle$ for RGGs and RRGs exhibits a monotonically increasing behavior similar to that observed previously in RSGs. It is remarkable that the sphere with $r=\sqrt{\pi}/(2\pi)$ (the rectangle with $a>1$) shows the smallest (largest) value of $\left \langle C \right \rangle$, when $s \approx s_c$. Indeed, the larger the length of the perimeter (i.e. the larger the value of $a$), the larger the value of $\left \langle C \right \rangle$ at $s \approx s_c$. It is also worth mentioning that, in the case of the rectangle with $a=10$, $\left \langle C \right \rangle \approx 3/4$ at $s \approx s_c$, which coincides with the exact value for one-dimensional RGGs according to Ref.~\cite{Dall02}. On the other hand,  surprisingly, when $s \approx s_c$, the analytical average clustering coefficient for two-dimensional RGGs (according to Ref.~\cite{Dall02}) is closer to the one observed for RSGs than the result obtained for RGGs. The reason why this happens is that analytical calculation in Ref.~\cite{Dall02} does not consider the boundary effects, whereas our numerical estimation for RGGs does. Finally for a wider description regarding the influence of $a$ on the average clustering coefficient of RRGs the reader is referred to Ref. \cite{Estrada15a}.

\section{Conclusions}
\label{conclusions}

In this work, we have developed a new model of random geometric network
in which we embed the points on a sphere with radius $r$ instead
of on a square or a rectangle. As
usual in the random geometric networks, $N$ nodes are distributed uniformly and independently in a finite surface, and then two points are connected by an edge if the shortest distance between them, measured along the surface, is less than or equal to a given threshold, the connection radius $s$.

By setting $r=\sqrt{\pi}/(2\pi)$, we have shown analytically and numerically the differences between the topological properties of RSGs and those of RGGs and RRGs. In this respect
we have obtained analytical expressions or bounds for the
average degree, degree distribution, connectivity, average path
length, diameter and clustering coefficient of RSGs. In general, these
properties are mainly produced by the lack of borders of the RSGs and depend on the connection radius $s$, as well as on the
radius of the sphere, $r$.

The introduction of RSGs opens new possibilities for studying spatially embedded random graphs. For instance, the dynamics of cross-borders spreading processes that take place on a planetary scale (such as the global spreading of diseases, rumors or computer virus, among others) can be addressed easily with this new tool. On the other hand, since the transition to full connectivity on random geometric networks is strongly influenced by the details of the boundary, 
the generalization of the RSG model to other curved finite surfaces with novel borders can be of great interest.


\begin{acknowledgments}
The author would like to thank the anonymous referee for valuable suggestions that helped to improve this paper.
\end{acknowledgments}



\section*{APPENDIX: Dependence of the expected average degree on the connection radius for PB-RRG\MakeLowercase{s}.}
\setcounter{subsection}{0}
\label{appA}

\begin{figure}[h!]
\centering
\subfloat[]{\label{int_grande}\includegraphics[width=0.5\textwidth]{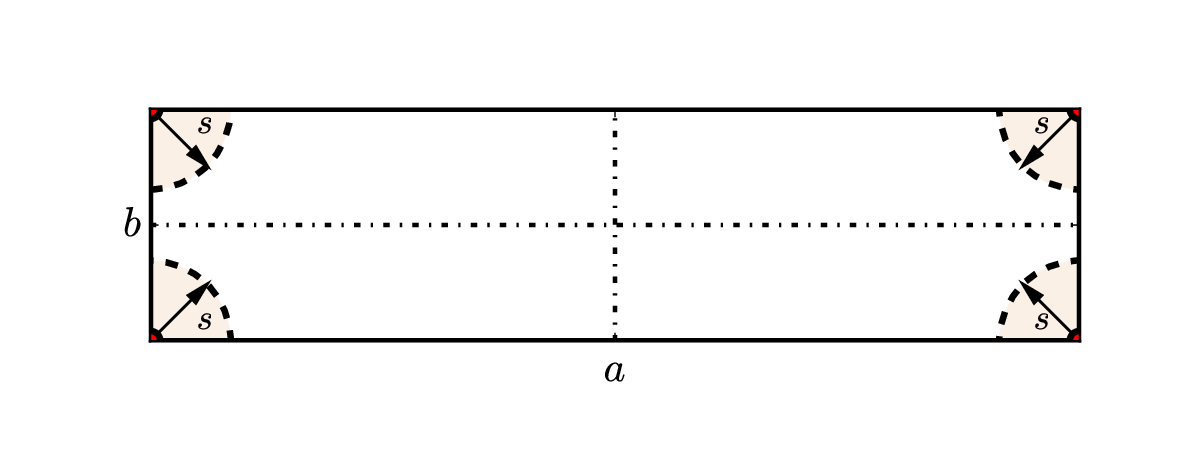}}

\medskip

\subfloat[]{\label{int_peq}\includegraphics[width=0.5\textwidth]{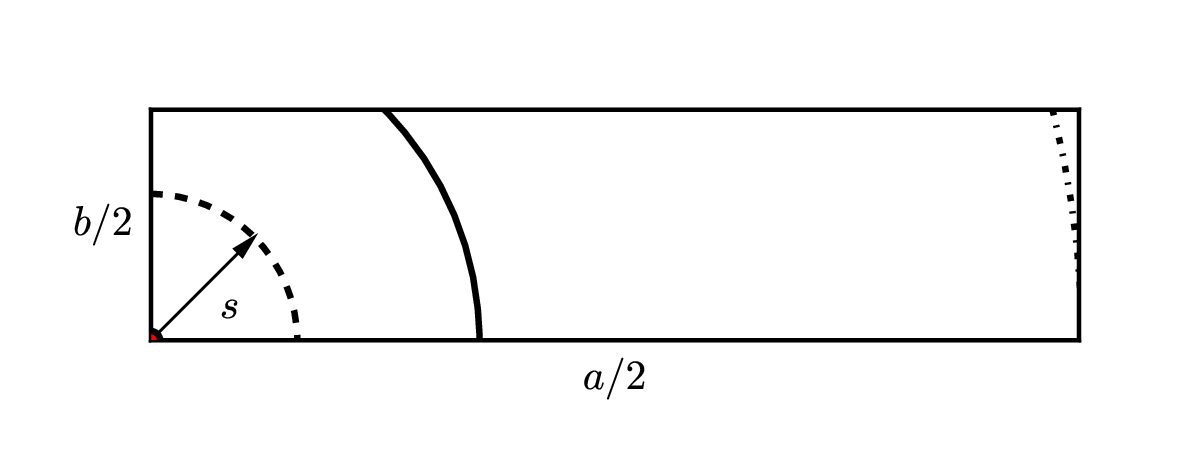}}
\caption{(a) Area within radius $s$ of a point which lies in the rectangle with periodic boundaries, $f_i^{PB} (s,a)$ (hatched area). (b) Illustration of three different quarter circles in the quarter rectangle, corresponding to $0\leq s \leq \frac{1}{2}b$ (dashed line), $\frac{1}{2}b\leq s \leq \frac{1}{2}a$ (continous line), and $\frac{1}{2}a\leq s \leq \sqrt{\frac{1}{4}a^2+\frac{1}{4}b^2}$ (dash-dotted line).}
\label{fig_integration}
\end{figure}

Let us consider a PB-RRG with $N$ nodes embedded in a rectangle with sides of lengths $a$ and $b$ (where $b \leq a$ and $b=1/a$), and let us suppose that, for a given node $i$, there are $N-1$ nodes distributed in the rest of the surface. Since the nodes are uniformly and independently distributed, the expected degree of a node $i$ can be expressed as

\begin{eqnarray}
\mathbf{E}\left [ k_i (s) \right ] = (N-1)\frac{f_i^{PB} (s,a)}{ab}= (N-1)f_i^{PB} (s,a),
\end{eqnarray}

\noindent where $f_i^{PB} (s,a)$ is the area within radius $s$ of a point which lies in the rectangle with periodic boundaries (see Fig.~\subref*{int_grande}). To calculate $f_i^{PB} (s,a)$, we determine the area of the quarter circle inside the quarter rectangle (see Fig.~\subref*{int_peq}), and multiply it by 4. Note that three different types of quarter circles can be used, when $s \leq \sqrt{a^2/4+b^2/4}$. In this way, we obtain Eq.~\ref{connection_area_rectangulo_PB}.

Considering that for a given value of $s$ all the points in $f_i^{PB} (s,a)$ are within the rectangle, it is possible to see that $f_i^{PB} (s,a)=f_j^{PB} (s,a)=f^{PB} (s,a)$ for $i\neq j$. Therefore, the expected degree of each node is the same and, consequently, its value is equal to the expected average degree of the PB-RRGs (see Eq.~\ref{demo_deg_PB}).

\begin{eqnarray}
\mathbf{E}\left [ \bar{k}(s) \right ]=\frac{\int_{S^2} \mathbf{E}\left [ k_i(s) \right ] \mathrm{d} S}{ab}=\mathbf{E}\left [  k_i(s) \right ]\frac{\int_{S^2} \mathrm{d}S}{ab}=\mathbf{E}\left [  k_i(s) \right ]=(N-1) f^{PB}(s,a).
\label{demo_deg_PB}
\end{eqnarray}

\noindent Thus, we obtain Eq.~\ref{eq_grado_medio_rectangulo_PB}.



\end{document}